 \documentclass[12pt]{article}
\usepackage{graphicx}
\usepackage{authblk}
\usepackage{snapshot}
\usepackage{amssymb}
\usepackage{lineno}
\usepackage[utf8]{inputenc}
\usepackage{amsmath}
\usepackage{graphicx}
\usepackage{multicol}
\usepackage{algorithmic}
\usepackage{subfigure}
\usepackage{program}
\usepackage[numbers,sort&compress]{natbib}

\title{Information dynamics algorithm for detecting communities in networks}
\author[1]{Emanuele Massaro}
\author[2]{Franco Bagnoli}
\author[3]{Andrea Guazzini}
\author[4]{Pietro Li\'o}
\affil[1]{Dept. of Informatic and Systems, Via S. Marta,3, Firenze, Italy}
\affil[2]{Dept. of Energy and CSDC, Via S. Marta,3, Firenze, Italy }
\affil[3] {Institute for Informatics and Telematics (IIT), National Research Council (CNR), via G. Moruzzi, 1 56124 Pisa, Italy}
\affil[4]{Computer Laboratory, University of Cambridge, 15 J.J. Thompson Avenue, Cambridge, CB30FD, UK}

\date{November 1, 2001}

\begin{document}
\maketitle
\begin{abstract}
The problem of community detection is relevant in many scientific disciplines, from social science to statistical physics. Given the impact of community detection in many areas, such as psychology and social sciences, we have addressed the issue of modifying existing well performing algorithms by incorporating elements of the domain application fields, i.e. domain-inspired. We have focused on a psychology and social network - inspired approach which may be useful for further strengthening the link between social network studies and mathematics of community detection.
Here we introduce a community-detection algorithm derived from the van Dongen's Markov Cluster algorithm (MCL) method~\cite{MCL} by considering networks' nodes as agents capable to take decisions. In this framework we have introduced a  memory factor to mimic a typical human behavior such as the \emph{oblivion effect}. The  method is based on information diffusion  and it includes a non-linear processing phase. We test our method on  two classical community benchmark and on computer generated networks with known community structure.  Our approach has three important features: the capacity of detecting overlapping communities, the capability of identifying communities from an individual point of view and the fine tuning the community detectability with respect to prior knowledge of the data. Finally we discuss how to use a Shannon entropy measure for parameter estimation in complex networks.
\end{abstract}

\section{Introduction}
Detecting communities is a task of great importance in many disciplines, namely sociology, biology and computer science~\cite{Waaserman,Scott,Mendes,Strogatz,Albert},  where systems are often represented as graphs. Community detection is also linked to clustering of data: many clustering methods establish links among representative points that are nearer than a given threshold, and then proceed in identifying communities on the resulting graphs~\cite{clustering1,domany}. 
Given a graph, in a broad sense, a community is a group of vertices ``more linked'' than between the group and the rest of the graph. This is clearly a poor definition, and indeed, on a connected graph, there is not a clear distinction between a community and the rest of the graph. In general, there is a continuum of nested communities whose boundaries are somewhat arbitrary: the structure of communities can be seen as a hierarchical dendogram~\cite{Newman}. 

In general, community detection algorithms rely on global  quantities like betweenness, centrality, etc.~\cite{Newman,communities} and most algorithms require the graph to be completely known. This constraint is problematic for networks like the World Wide Web, which for all practical purposes is too large and too dynamic to ever be fully known.

Moreover in complex networks, and in particular in social networks, it is very difficult to give a clear definition of community: it is caused by the fact that nodes often results in overlapping communities because they belong to more than one cluster or module or community. The problem of overlapping communities was discussed in~\cite{Palla} and recently a solution to it were presented in ~\cite{Lanc}. For instance people usually belong to different communities at the same time, depending on their families, friends, colleagues, etc.: so each people, making a \emph{subjective} community detection algorithm, has its own vision of communities in his social environment. 

In social networks, the definition of a community could be linked to the human capability of information processing, particularly the poor evaluation of probabilities. When faced with insufficient data or insufficient time for a rational processing, we humans have developed algorithms, denoted heuristics, that allows to take decisions in these situations. The modern approach to the study of cognitive heuristics defines them as those \textit{strategies that prevent one from finding out or discovering correct answers to problems that are assumed to be in the domain of probability theory}. The ratio of a cognitive algorithm for community detection is based on the fact that humans' networks are the results of the individual stategies of  single subjects; on the other hand they are presumably shaped and evolved by the social structures in which they live~\cite{Gigerenzer2002, Gigerenzer2011}.

The  paper is organized as follows: we start by describing a new algorithm for detecting communities in complex networks in Section~\ref{secondo}. Considering psychological notions as mentioned above, we adopted local algorithm where an individual is simply modeled as a memory and a set of connections to other individuals. The ``learning''  (nonlinear) phase is modeled after competition in chemical/ecological world, where resources fighting each other in order not to fall into oblivion.  In Section~\ref{terzo} we describe the first algorithm in which  information about neighboring nodes is propagated and elaborated locally, but the  connections do not change. Here we want to emphasize not only the good efficiency of the algorithm in detecting community but also its capability to discover overlapping nodes and a sort of subjective vision of hierarchical levels of the network. Next, in Section~\ref{quinto} we give an interpretation of Shannon entropy of information as quality function for estimating models parameters. Finally we discuss our results and we propose future steps in Conclusions.

\section{Competion process}
\label{secondo}
\begin{figure}[t!]
\centerline{\includegraphics[width=.8\columnwidth]{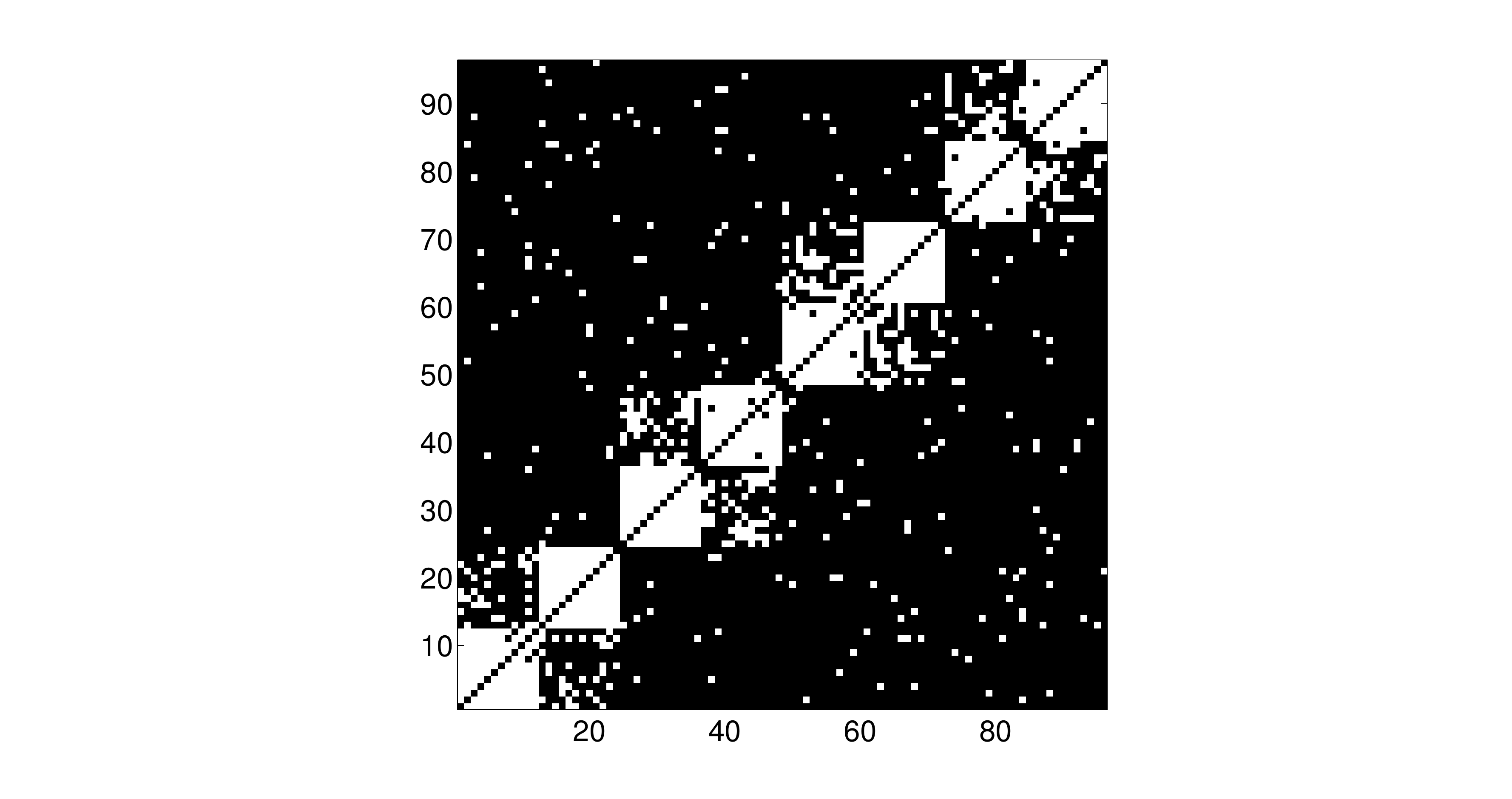}}
\caption{An example of adjacency matrix $A$. It is a three-level matrix composed by 4 blocks of 2 sub-communities of 8 nodes each. The link probability inside a sub-community is $0.98$,   in first level blocks is $0.3$ and among  blocks is $0.03$. White points indicate the presence of a link between the node $i$ and the node $j$, $A_{ij} = 1 $.}
\label{f.lbl}
\end{figure}

We consider $N$ individuals, labeled from $1$ to $N$. 
Let us denote by $A$ the adjacency matrix, $A_{ij} = 1 $ (0) indicates the presence (absence) of a link from site $j$ to site $i$; all links have the same weight (\figurename~\ref{f.lbl}). 
Each individual $i$ is characterized by a state vector $S_i$, representing his knowledge of the outer world. We interpret $S$ as a probability distribution, assuming that $S_i^{(k)}$ is the probability that individual $i$ belongs to the community $k$. Thus, $S_i^{(k)}$ is normalized on the index $k$. We shall denote by $S=S(t)$ the state of the all network at time $t$, with $S_{ik} = S_i^{(k)}$. We shall initialize the system by setting $S_{ij}(0) = \delta_{ij}$, where $\delta$ is the Kroneker delta, $\delta_{ij}=1$ if $i=j$ and zero otherwise. In other words, at time 0 each node only knows about itself.  

As mentioned the competition phase is modeled thinking to a chemical/ecological analogy. 
Our algorithms are based on the concept of \emph{diffusion and competitive interaction} in network structure introduced by Nicosia et al.~\cite{Nicosia09}. 

If two populations $x$ and $y$ are in competition for a given resource, their total abundance is limited~\cite{murray2002}.  After normalization, we can assume $x+y=1$, i.e., $x$ and $y$ are the frequency of the two species, and $y=1-x$. The reproductive step is given by $x' = f(x)$, which we assume to be represented by a power $x'=x^\alpha$. For instance, $\alpha=2$ models the birth of individuals of a new generation after binary encounters of individuals belonging to the old generation, with noneverlapping generations (eggs laying)~\cite{frei}.

After normalization we obtain:
\begin{equation}
 x' = \frac{x^\alpha}{x^\alpha + y^\alpha} = \frac{x^\alpha}{x^\alpha + (1-x)^\alpha}.
\end{equation}
Introducing $z=(1/x)-1$ ($0\le z< \infty$), we get the map
\begin{equation}  
\label{ref:second}
 z(t+1) = z^\alpha(t), 
\end{equation}
whose fixed points (for $\alpha > 1$) are 0 and $\infty$ (stable attractors) and $1$ (unstable), which separates the basins of the two attractors. Thus, the initial value of $x$, $x_0$, determines the asymptotic value, for $0\le x < 1/2$, $x(t\rightarrow\infty) = 0$, and for $1/2< x < 1$, $x(t\rightarrow\infty) = 1$. 

By extending to a larger number of components for a probability distribution $P_i$, the competition dynamics becomes
\begin{equation}
  P_{i}' = \frac{P_{i}^\alpha}{\sum_j P_{j}^\alpha},
  \label{ref:competizione}
\end{equation}
and the iteration of this mapping, for $\alpha>1$,   leads to a Kroneker delta, corresponding to the larger component. However, the alternation between information and competition can generate interesting behaviors.

\section{Information Dynamics Algorithm}
\label{terzo}
The dynamics of the network is given by an alternation of communication and elaboration phases. Communication is implemented as a simple diffusion process, with memory $m$. The memory parameter $m$ allows us to introduce some limitations in human cognition such as the mechanism of oblivion and the timing effects: the most recent information has more relevance than the previous one ~\cite{Tulving82,Forster84}. 

We assume that each individual spends the same amount of time in communication, so that people with more connections dedicate less time to each of them. Since the amount of available time is limited, we normalize the adjacency matrix on the columns (i.e., we assign at each link the inverse of the output degree of the incoming node), forming a Markov matrix $M$
\begin{equation}
  M_{ij} = \frac{A_{ij}}{\sum_k A_{kj}}.
\end{equation}
Note that in many mathematical texts the indices are inverted, so that the Markov matrices are normalized on the rows. We prefer the ``physics'' notation so that matrix multiplication with a probability distribution $P$ takes the usual form $P' = M P$. Then in the communication phase, the state of the system evolves as

\begin{equation}
  S(t+\frac{1}{2}) = m S(t) + (1-m) M S(t).
\end{equation}

As described in the Eq. \ref{ref:competizione}, the competition phase is modeled thinking to a competive interaction between the nodes in the network~\cite{Nicosia09}.

In this way the dynamic of the model is given by a sequence $S(t)\rightarrow S(t+\frac{1}{2})\rightarrow S(t+1)$:
\begin{equation}
\label{ref:bagnongen} 
 \begin{split}
   {S}_{ik}\left( t+\frac{1}{2}\right) &= m S_{ik}(t) + (1-m) \sum_j M_{ij} S_{jk}(t), \\ 
   S_{ik}(t+1) &= \frac{{{S}}_{ik}^{\alpha}(t+\frac{1}{2})}{\sum_j {{S}}_{ij}^{\alpha}(t+\frac{1}{2})}.
 \end{split}
\end{equation}

We assume that individuals' memory is large enough so that they can keep track of all information about all other individuals. In a real case, one should limit this memory and apply an input/output filtering. Individuals do not change their connectivity. 
For testing purposes we use three networks  and analyzing and discussing our model peculiarities.  The three case studies, of growing or different complexity, are: a simple artificial network used to show the typical output of our algorithm, the  Zachary \emph{karate club} network  \cite{Zachary} and the bottlenose dolphins network~\cite{dolphin}.

\subsection{Simple artificial network}

\begin{figure}[h!]
  \label{artificiale} 
  \begin{center}
      \includegraphics[width=0.5\columnwidth]{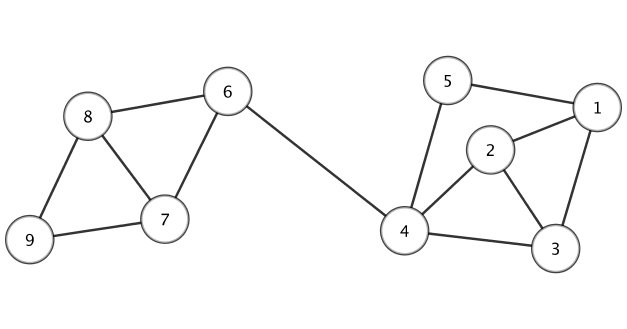}
  \end{center}
    \caption{\label{fig:artificiale} \emph{Simple artificial network} composed by of 9 nodes and 13 links divided in 2 communities. It is possible to identify two different communties: the first one composed by nodes 1-2-3-4-5 and the second one by 6-7-8-9.}
\end{figure}

In this first case study the algorithm face with a very simple task and converges to an optimal solution in few iterations and for a wide range of model's parameters $m$ and $\alpha$. Analyzing state matrix S(t), it is possible to identify two different communities marked by nodes 5 and 9.

\begin{tiny}
\begin{center}
S(T)=
$\begin{pmatrix}
0 & 0 & 0 &  0 & 0.9999 & 0 &  0 & 0 &  0.0001\\ 
0 & 0 & 0 &  0 & 0.9982 & 0 &  0 & 0 &  0.0018\\
0 & 0 & 0 &  0 & 0.9982 & 0 &  0 & 0 &  0.0018\\
0 & 0 & 0 &  0 & 0.9383 & 0 &  0 & 0 &  0.0617\\
0 & 0 & 0 &  0 & 0.9975 & 0 &  0 & 0 &  0.0025\\
0 & 0 & 0 &  0 & 0.1309 & 0 &  0 & 0 &  0.8691\\
0 & 0 & 0 &  0 & 0.0054 & 0 &  0 & 0 &  0.9946\\
0 & 0 & 0 &  0 & 0.0054 & 0 &  0 & 0 &  0.9946\\
0 & 0 & 0 &  0 & 0.0061 & 0 &  0 & 0 &  0.9939\\
\end{pmatrix}$ 
\end{center}
\end{tiny}

\begin{figure}[h!]
\centering
\subfigure[]
{\includegraphics[width=6cm]{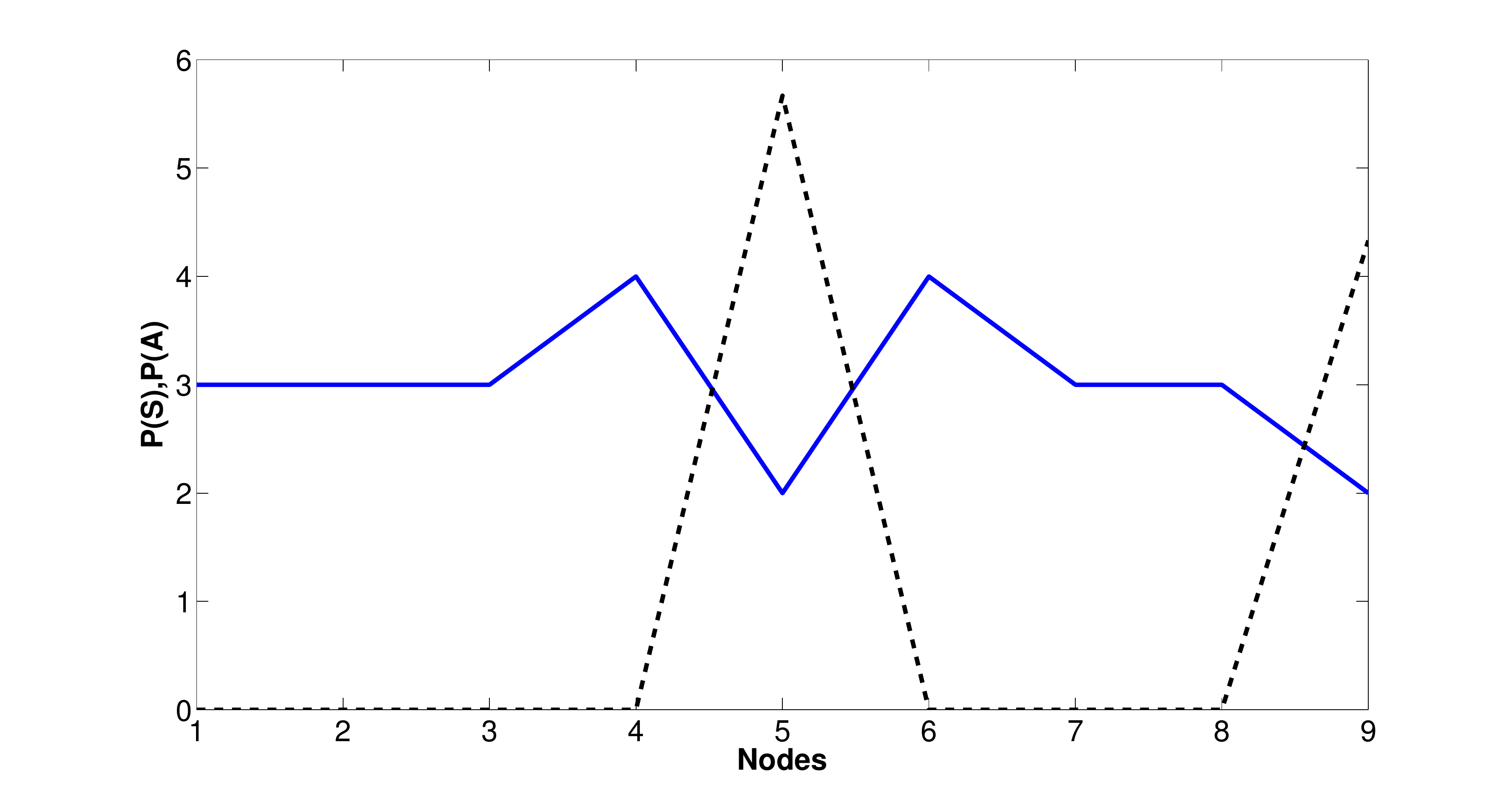}}
\hspace{5mm}
\subfigure[]
{\includegraphics[width=6cm]{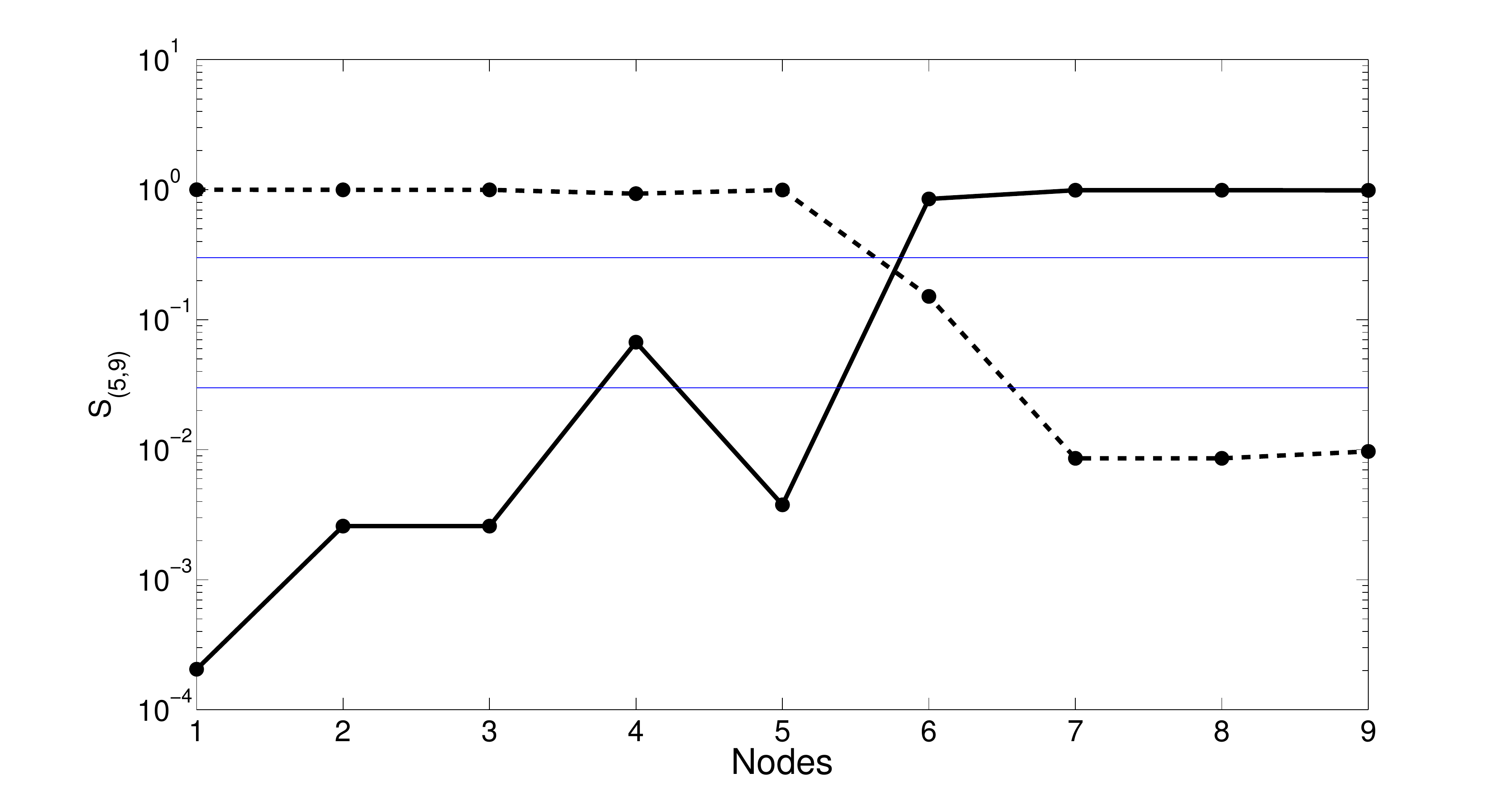}}
\caption{\label{fig:cfart1}(a) On the x-axis of both figures there are is number of nodes. On the y-axis: the cumulative distribution  $P^{(S)}$ (dashed black line, $P_j^{(S)}=\sum_i S_{ij}$, multiplied by five) and $P^{(A)}$ (blue line, $P_j^{(A)}=\sum_i A_{ij}$, connectivity). The information propagation algorithm identifies communities by leaves (nodes 5 and 9 with lower connectivity) with $m=0.3$ and $\alpha = 1.4$. (b) The value of state vectors, at the final asymptotic time, of node 5 (dashed black line) and node 9 (black line). We can observe upper values indentifying communities: the first one composed by nodes 1-2-3-4-5 and the second one by nodes 6-7-8-9. The algorithm is capable also to detect the \emph{communication nodes} 4 and 6 between the blue lines.  In this way we can indentify the overlap between the communities and also define a sort of \emph{objective vision} of nodes. It is clear that the upper nodes know very well which is their community as well as nodes 4 and 6 that know that they are in a middle state between two communities.}
\end{figure}

In \figurename~\ref{fig:cfart1}(b) it is possible to identify two different communties highlighted by upper values in the graph. The first community is composed by node 1-2-3-4-5 and the second one by 6-7-8-9. Our algorithm is capable also to detect overlapping nodes (4 and 6) as "middle" values between blue lines. In this way each node knows exactly its role in the network.

\subsection{Zachary "Karate Club"}
\begin{figure}[h!]
\centering
\subfigure[]
{\includegraphics[width=6cm]{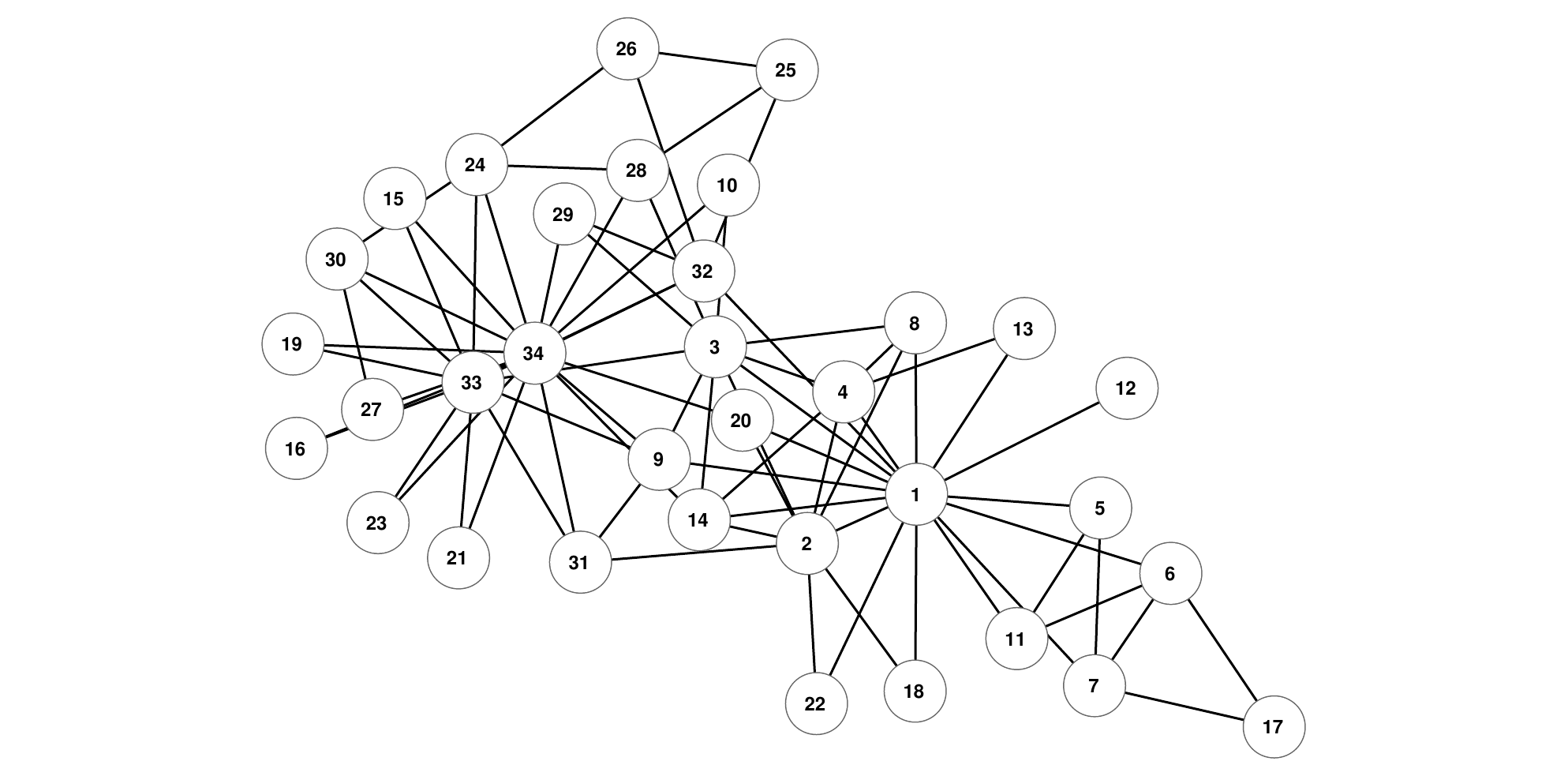}}
\hspace{5mm}
\subfigure[]
{\includegraphics[width=6cm]{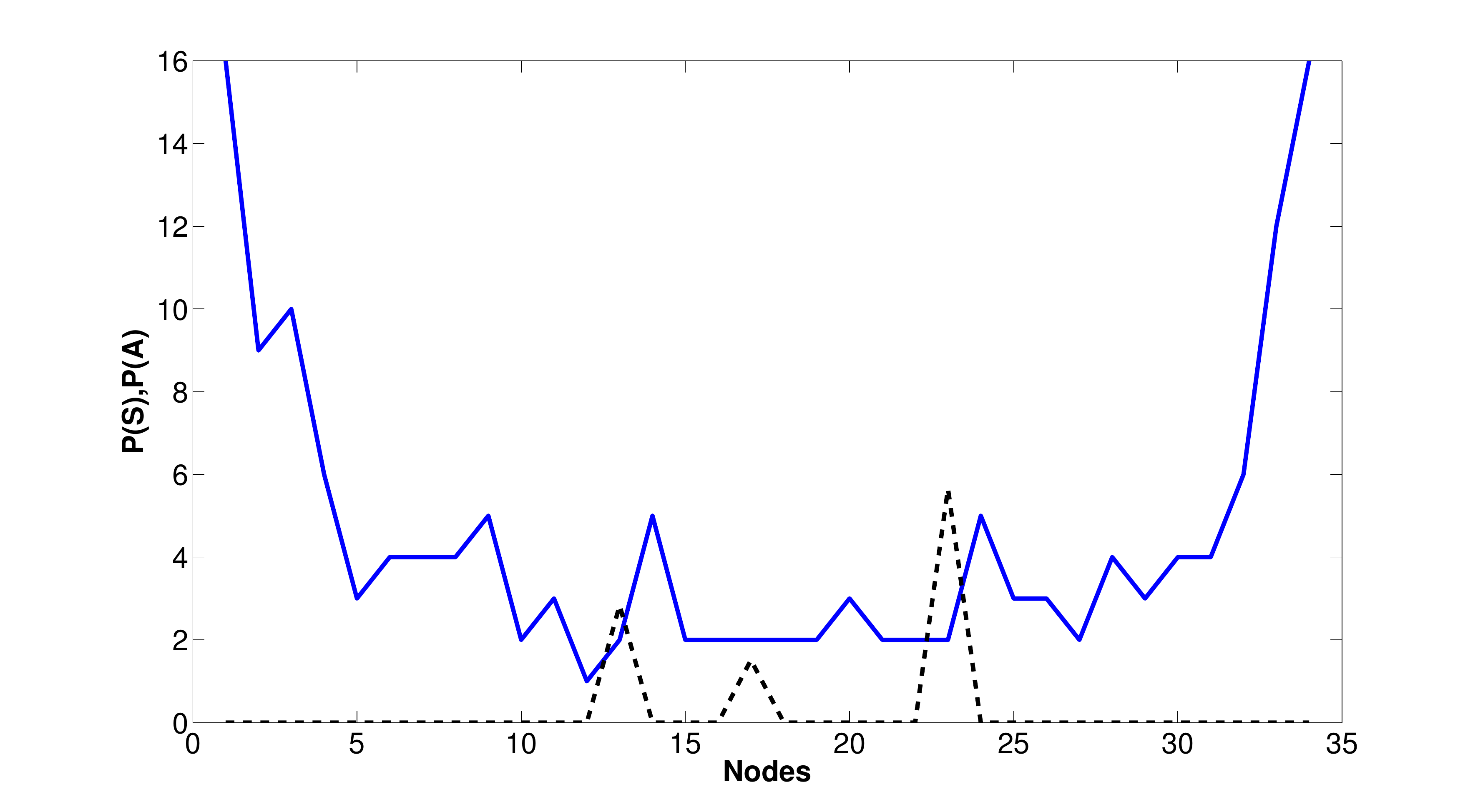}}
\caption{\label{fig:karate}(a) Zachary's karate club network. (b) On the x-axis of the figure there is the number of nodes. On the y-axis: the connectivity (blu line) $P^{(A)}$ and the cumulative distribution (dashed black line)  $P^{(S)}$ are reported at final asymptotic time with $m=0.2$ e $\alpha=1.4$. The $P^{(S)}$ reveals three underlying substructures labeled by nodes 13,17 and 23. }
\end{figure}

The second test case is a typical network literature example: the network proposed by Zachary in the 1977 , and known as \textit{"karate club"} \cite{Zachary}. Although this network (\figurename~\ref{fig:karate}(a)) is rather small, our algorithm shows interesting results. With $m=0.25$ and $\alpha=1.4$ the algorithm has detected three communities in few steps as described in the \figurename~\ref{fig:karate}(b).

\newpage
\begin{figure}[h!]
\centering
\subfigure[]
{\includegraphics[width=6cm]{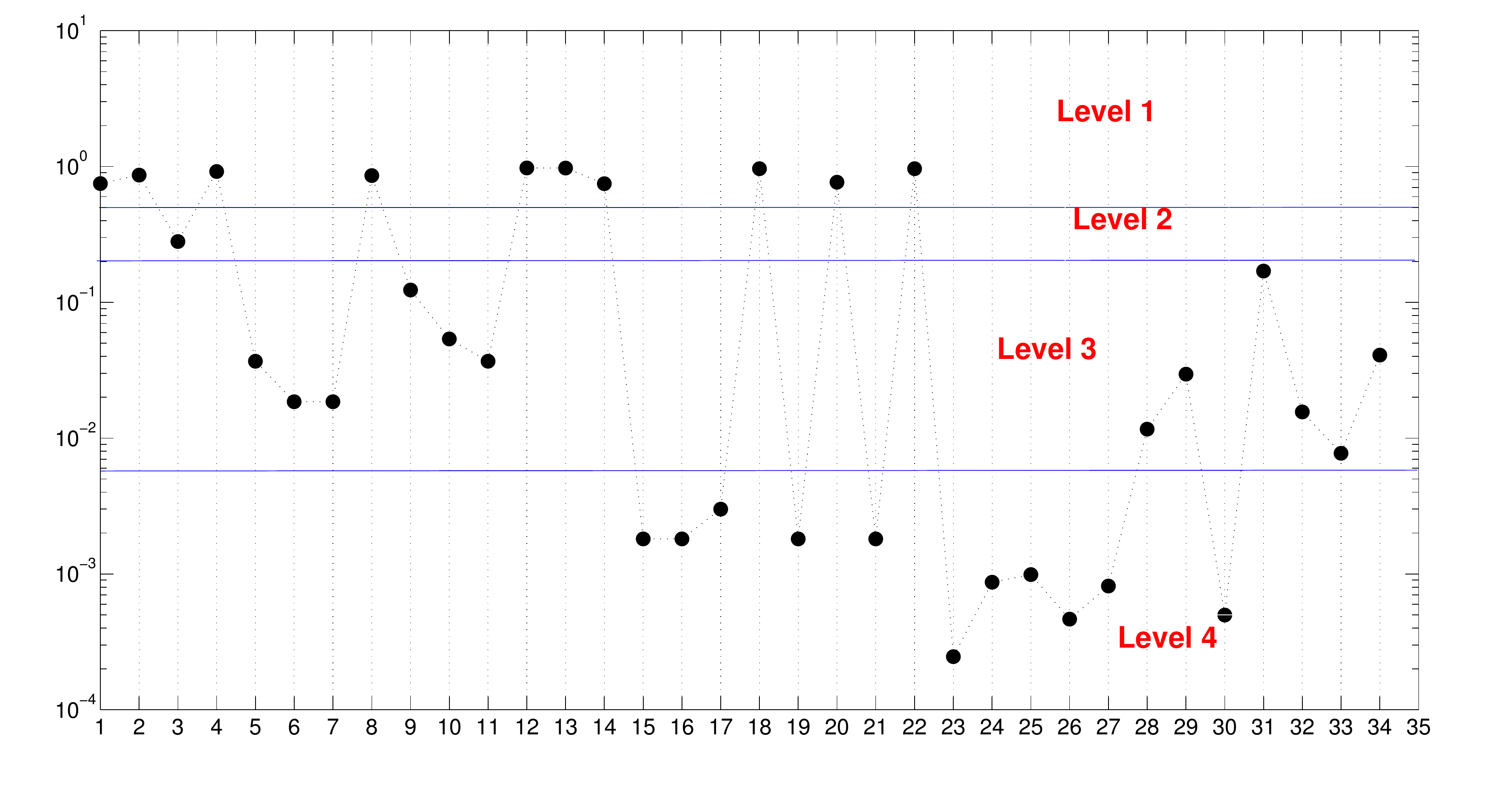}}
\hspace{5mm}
\subfigure[]
{\includegraphics[width=6cm]{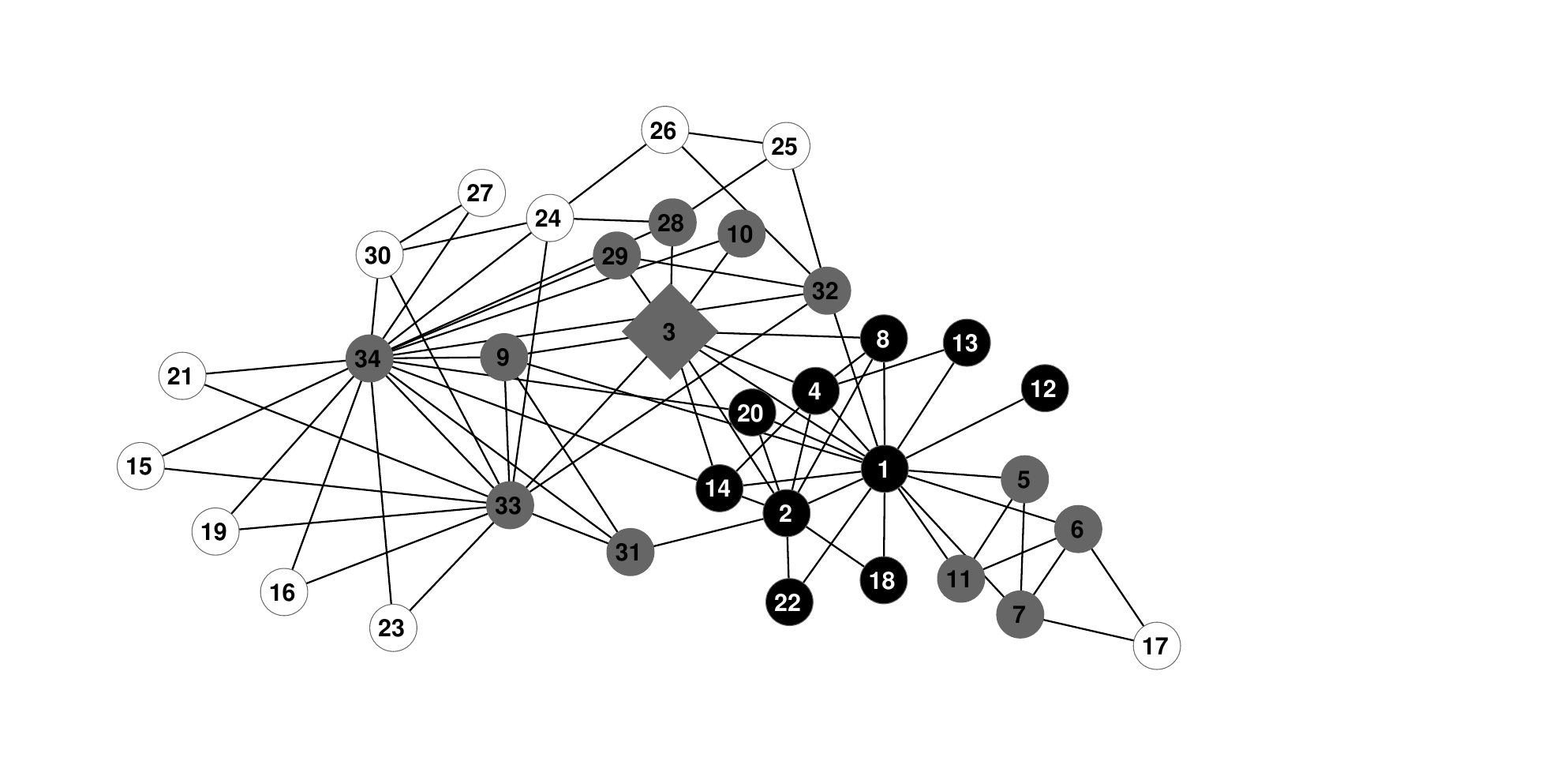}}
\caption{\label{fig:karate13}(a) Values of state vector 13: the community is defined by black nodes in \figurename~\ref{fig:karate13}(b) corresponding to upper values in the Level 1. As well as in the simple artificial network network the algorithm has detected not only the rigth community but also the overlapping node corresponding to big gray diamond in \figurename~\ref{fig:karate13}(b), in the Level 2. The Level 3  and the Level 4 (respectively gray and white nodes in \figurename~\ref{fig:karate13}(b)) correspond to the different level of knowledge the others. In the Level 3 it is possible to find "direct" contacts while in the fourth level it is possible to detect "friends of my friends".}
\end{figure}

\begin{figure}[h!]
\centering
\subfigure[]
{\includegraphics[width=6cm]{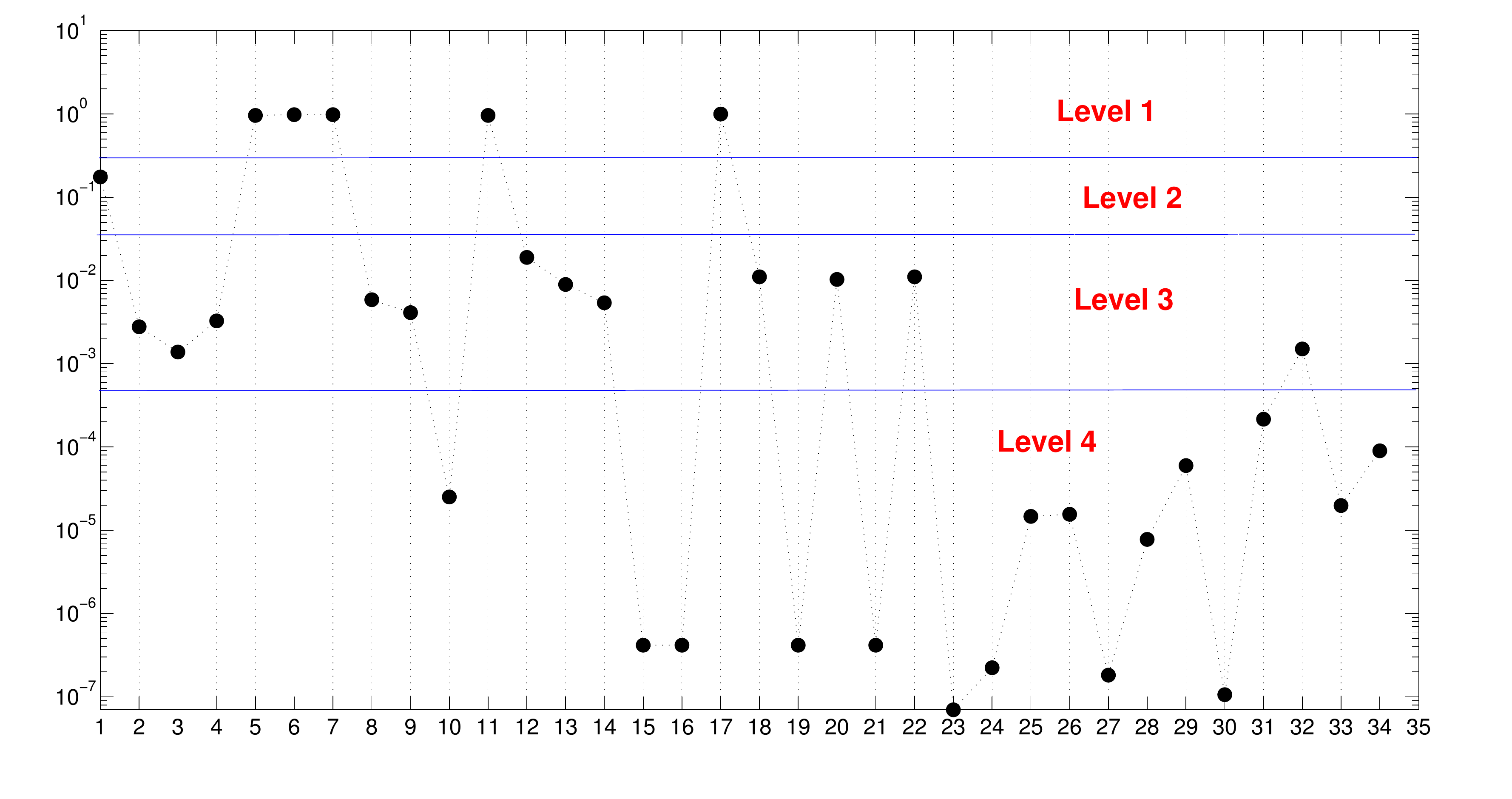}}
\hspace{5mm}
\subfigure[]
{\includegraphics[width=6cm]{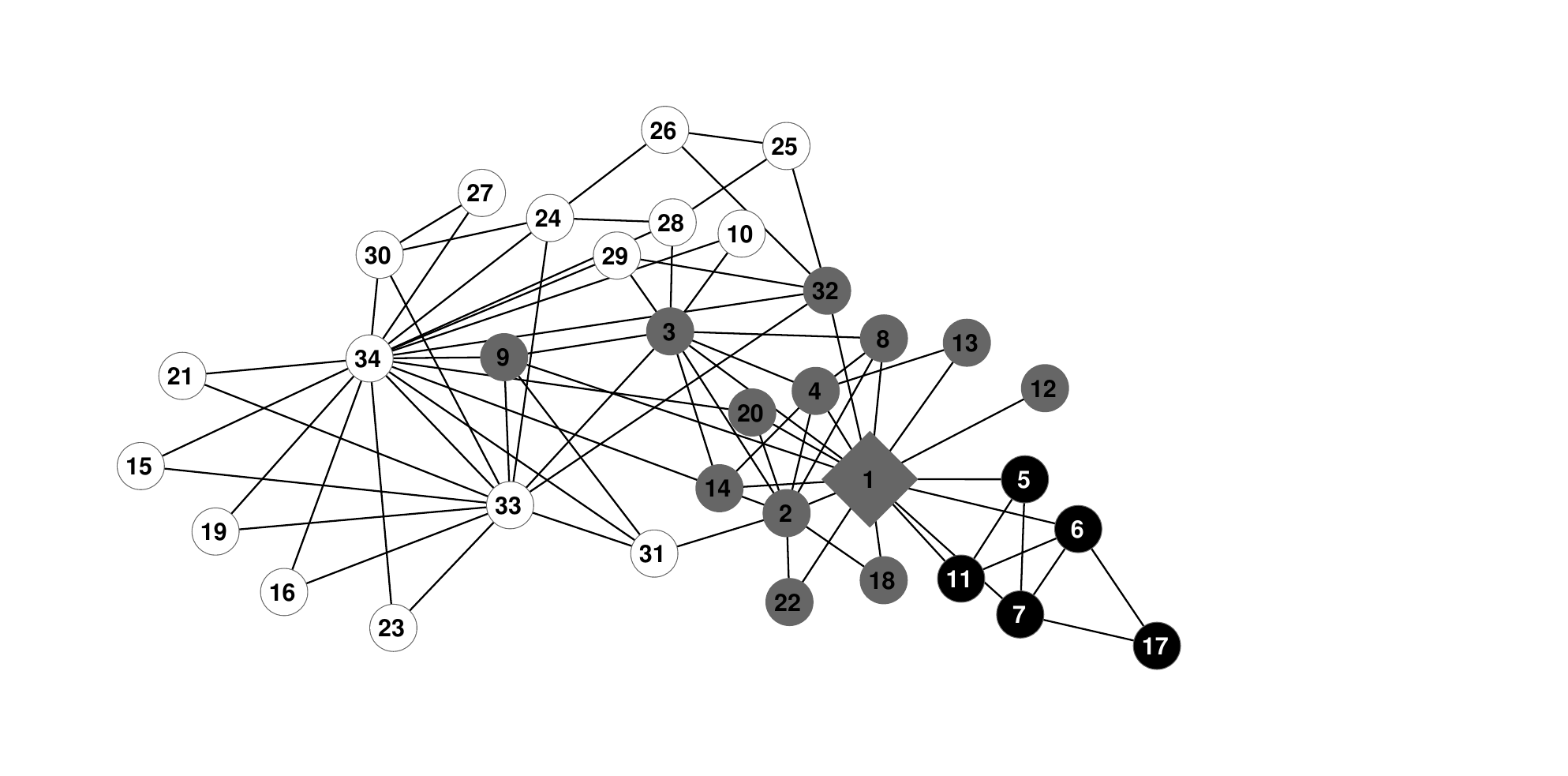}}
\caption{\label{fig:karate17}(a) Values of state vector 17: the community is defined by black nodes in \figurename~\ref{fig:karate17}(b) corresponding to upper values in the Level 1. For the description see the caption of the \figurename~\ref{fig:karate13}.}
\end{figure}
\newpage
\begin{figure}[h!]
\centering
\subfigure[]
{\includegraphics[width=6cm]{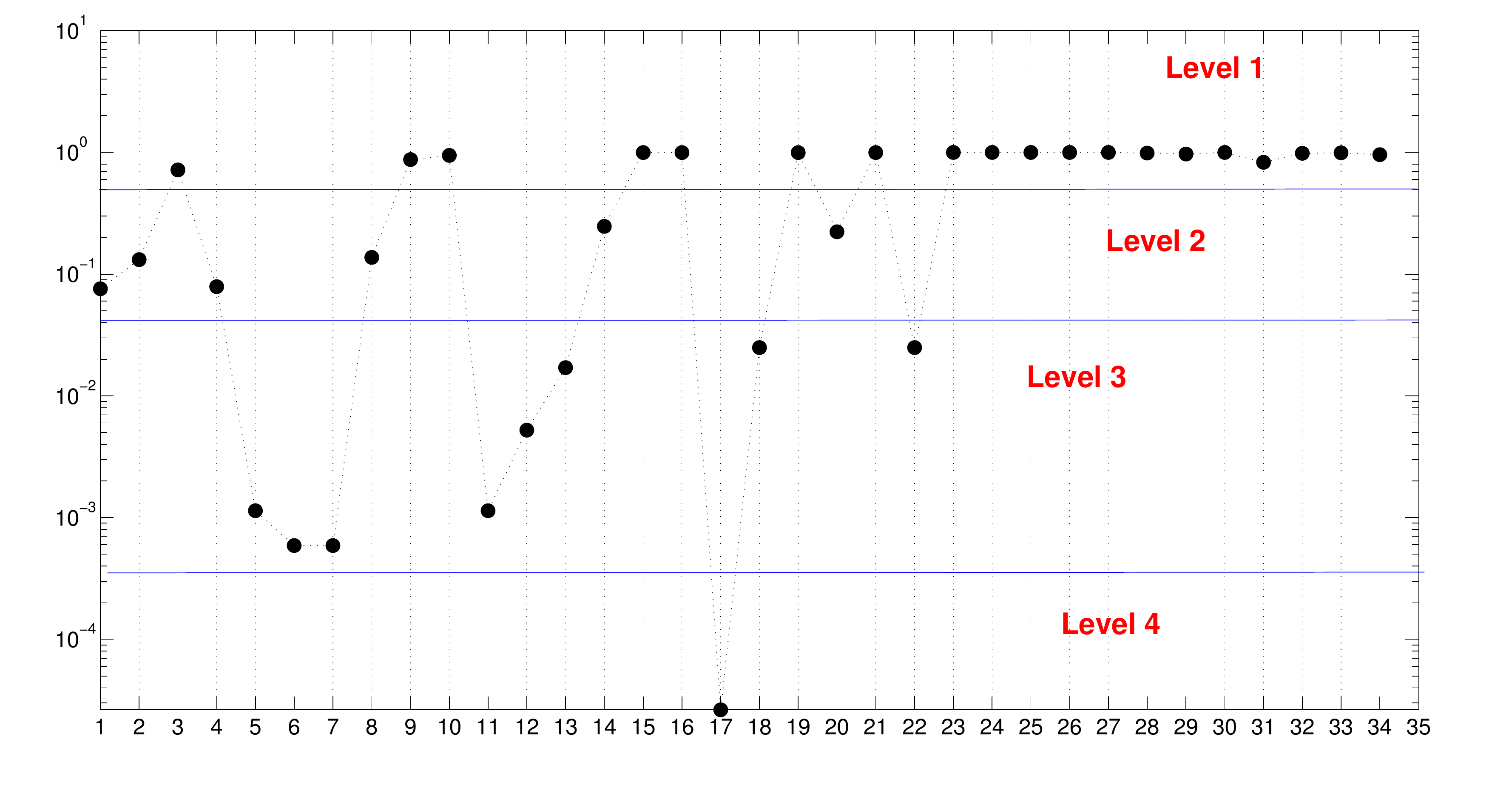}}
\hspace{5mm}
\subfigure[]
{\includegraphics[width=6cm]{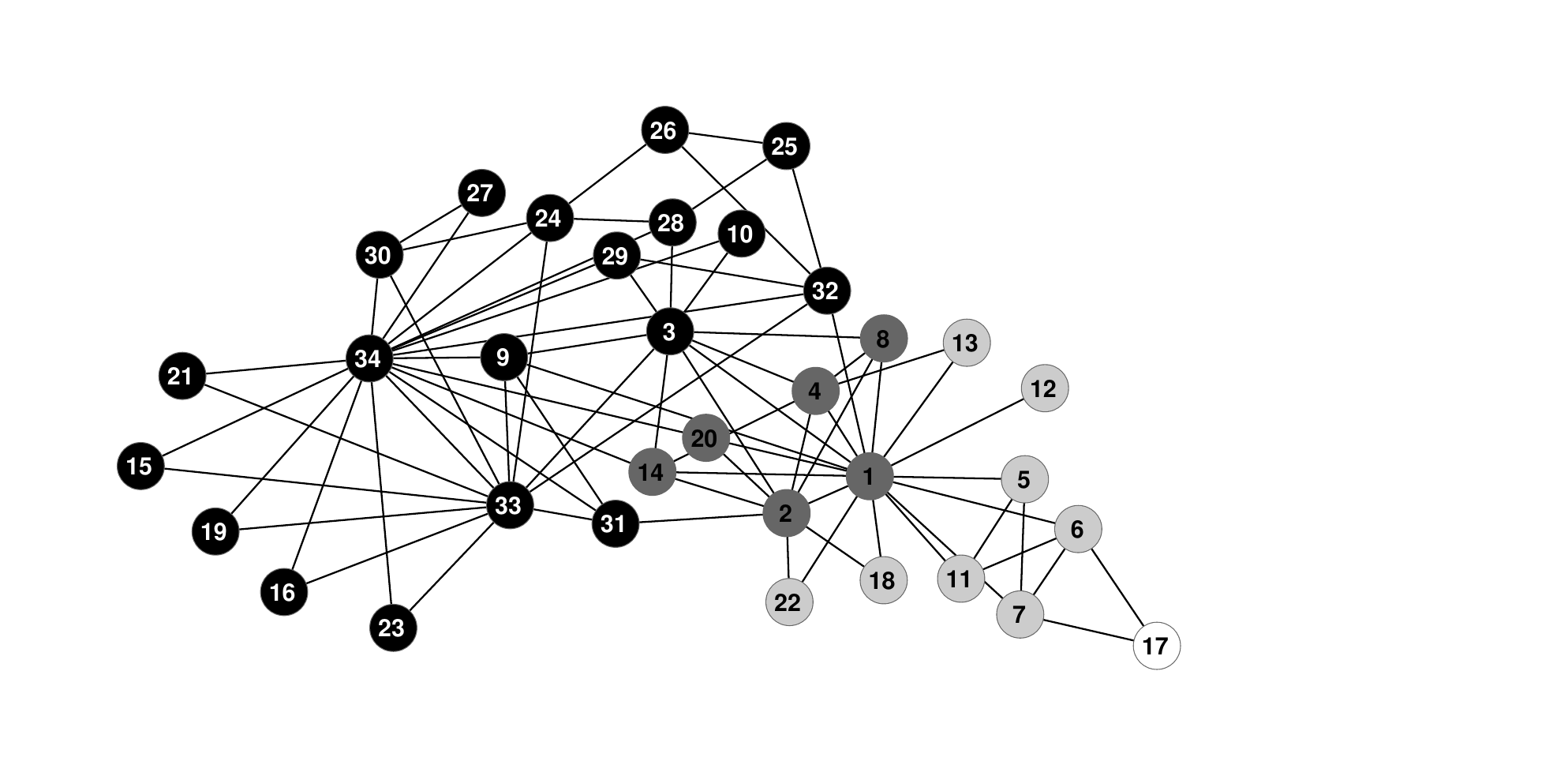}}
\caption{\label{fig:karate23}(Values of state vector 17: the community is defined by black nodes in \figurename~\ref{fig:karate23}(b) corresponding to upper values in the Level 1. In this case we haven't found overlapping but is also possible to detect a sort of scale of friendship inside the network labeled by the 4 levels in \figurename~\ref{fig:karate23}(a) and by black-to-white color scale in \figurename~\ref{fig:karate23}(b) .}
\end{figure}

\begin{figure}[h!]
  \label{artificiale} 
  \begin{center}
      \includegraphics[width=1\columnwidth]{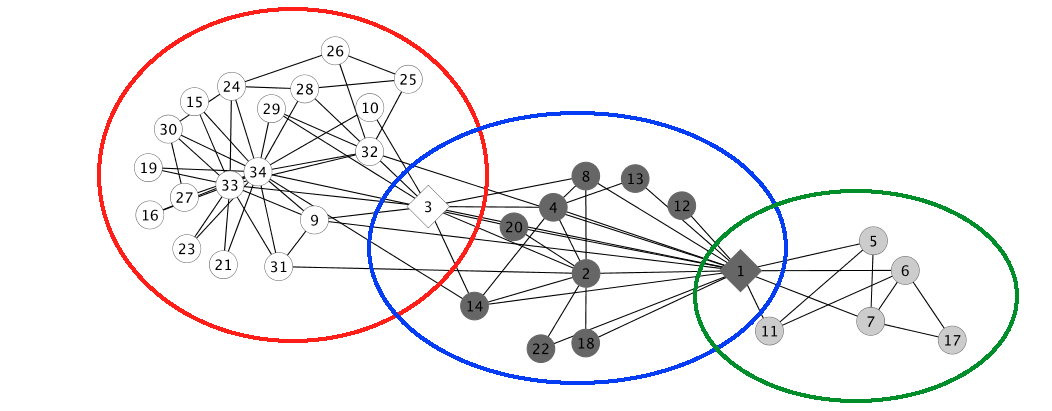}\\
  \end{center}
    \caption{\label{fig:karatecomm} Communities detected by our algorithm labeled by dark gray (shown in \figurename~\ref{fig:karate13}), light gray (shown in \figurename~\ref{fig:karate17}) and white nodes (shown in  \figurename~\ref{fig:karate23}). On the other hand  three circles rapresnt the overlap between communities because of the role of node 3 and 9 as explained in \figurename~\ref{fig:karate13} and \figurename~\ref{fig:karate17}. }
\end{figure}

\subsection{Bottlenose dolphin Network}
The third case study concerns  a community network of dolphins. The network we study was constructed from observations of a community of 62 bottlenose dolphins over a period of seven years from 1994 to 2001~\cite{dolphin}.
\begin{figure}[h!]
\centering
\subfigure[]
{\includegraphics[width=6cm]{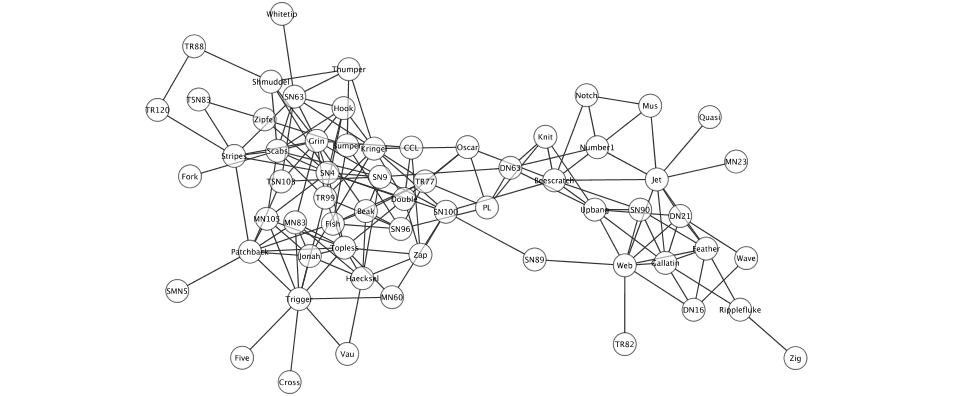}}
\hspace{5mm}
\subfigure[]
{\includegraphics[width=6cm]{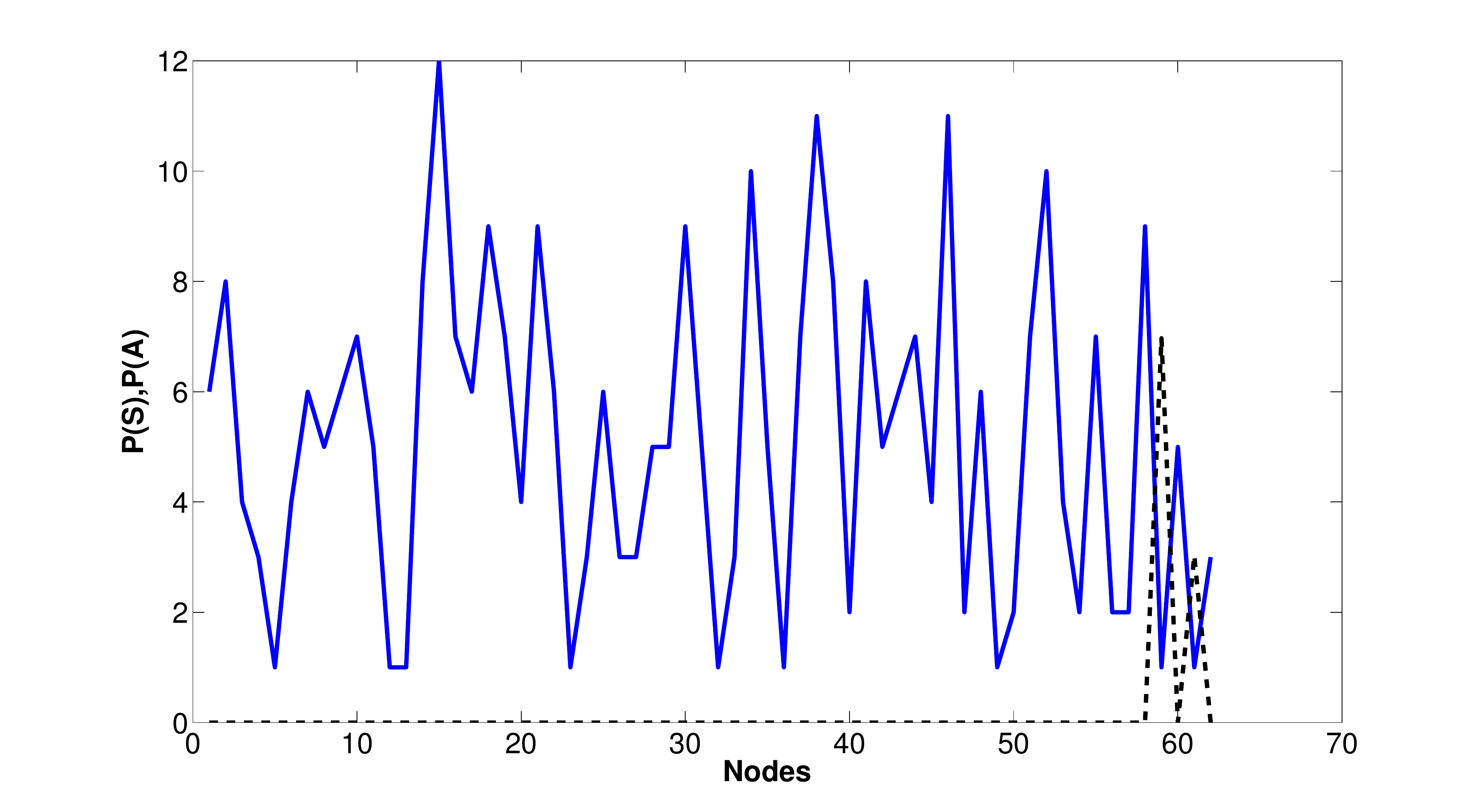}}
\caption{\label{fig:dolphin}(a) Bottlenose dolphin network. This network has a size of 62 nodes and from direct observation it is known that it has two communities.(b) On the x-axis of the figure there is the number of nodes. On the y-axis: the connectivity (blu line) $P^{(A)}$ and the cumulative distribution (dashed black line)  $P^{(S)}$ are reported at final asymptotic time with $m=0.5$ e $\alpha=1.03$. The $P^{(S)}$ reveals two underlying substructures labeled by nodes 59 and 61}
\end{figure}

\begin{figure}[h!]
\centering
\subfigure[]
{\includegraphics[width=6cm]{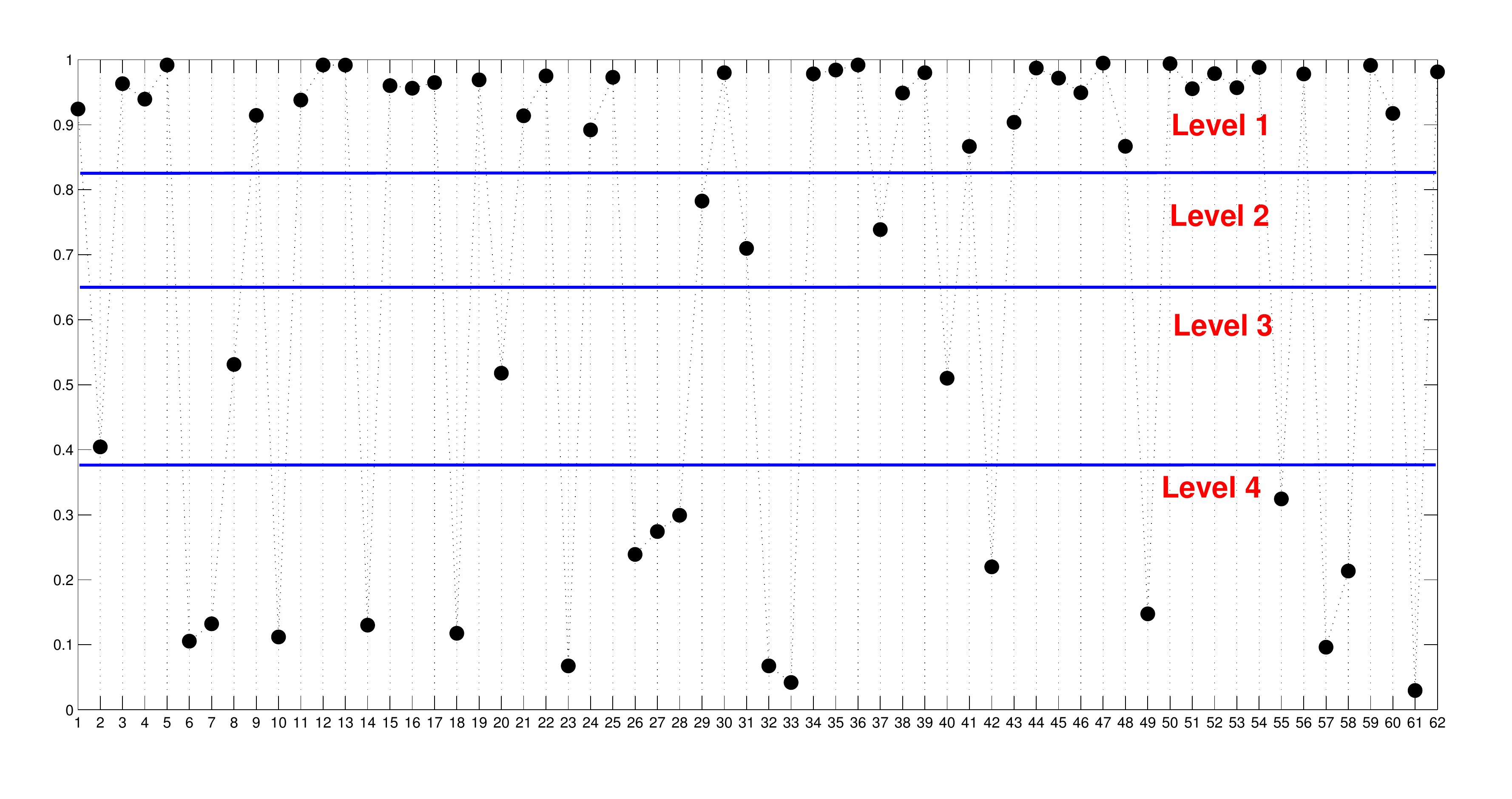}}
\hspace{5mm}
\subfigure[]
{\includegraphics[width=6cm]{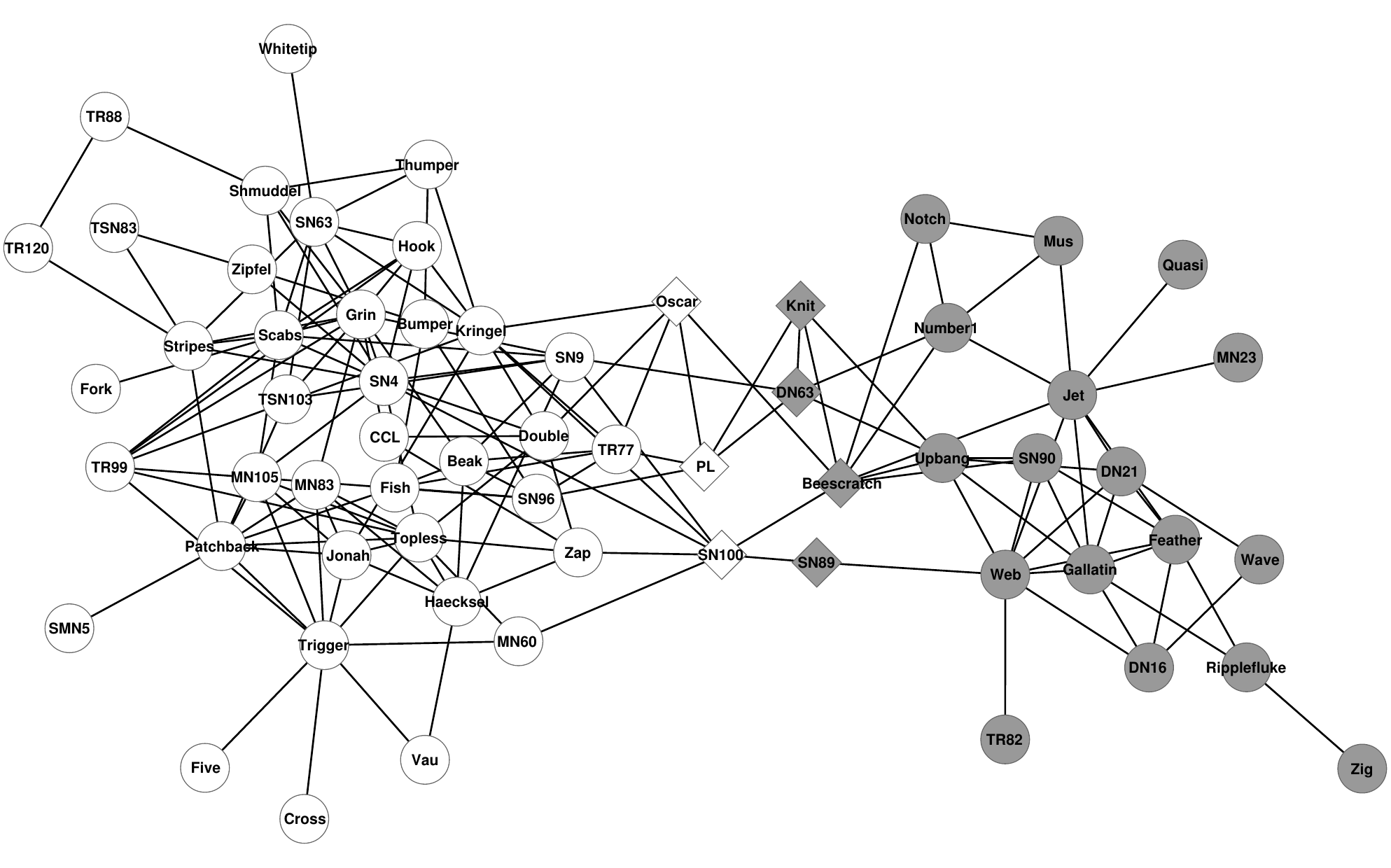}}
\caption{\label{fig:dolphin1}(a) Values of state vector number 59: the four levels indicate respectively nodes of community (white nodes), overlapping nodes (white diamond nodes), closer nodes of the other community(gray diamonds) and finally nodes of the other community (gray nodes).(b) Communities detected by our algorithm.}
\end{figure}

\begin{figure}[h!]
\centering
\subfigure[]
{\includegraphics[width=6cm]{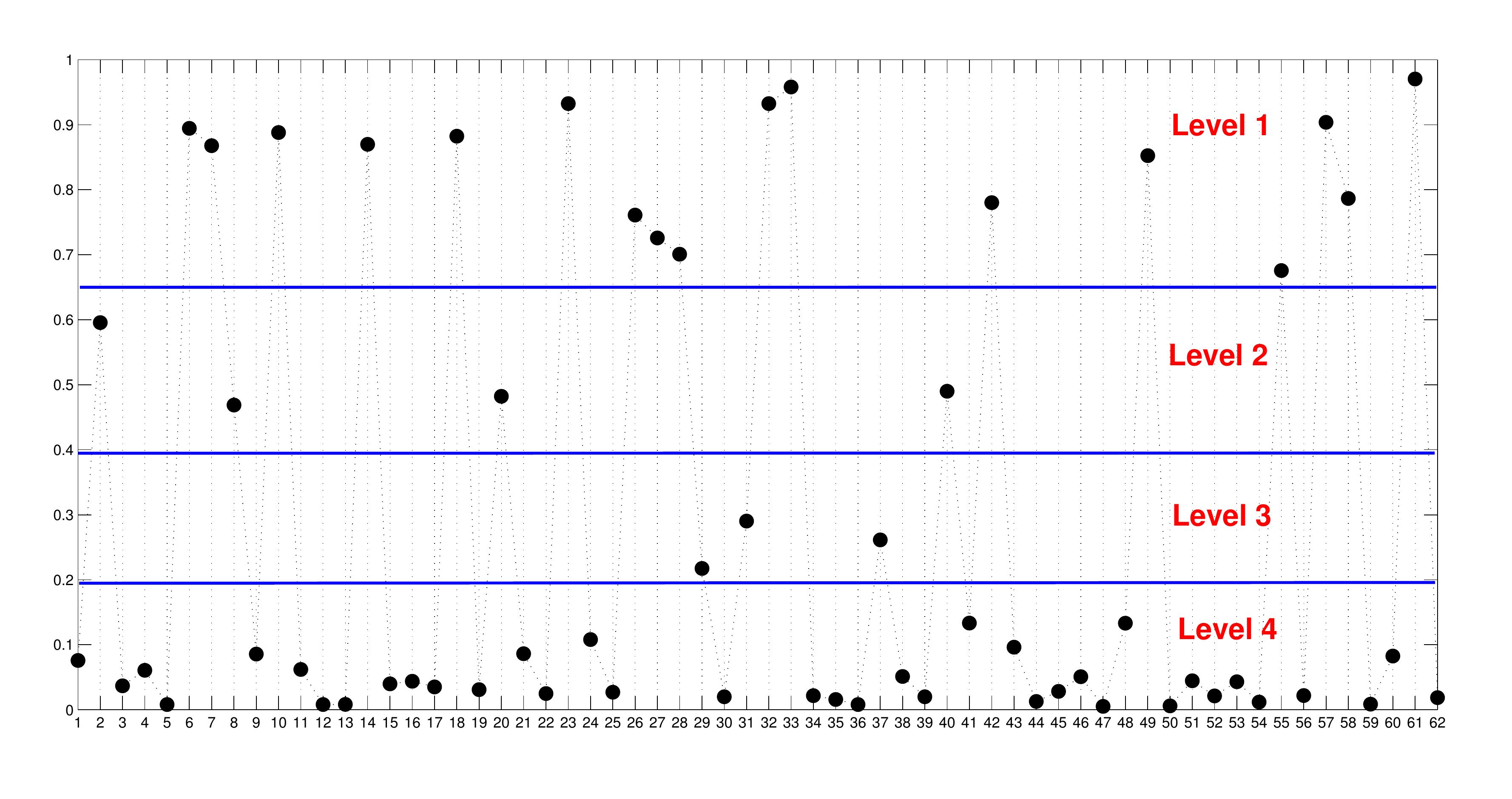}}
\hspace{5mm}
\subfigure[]
{\includegraphics[width=6cm]{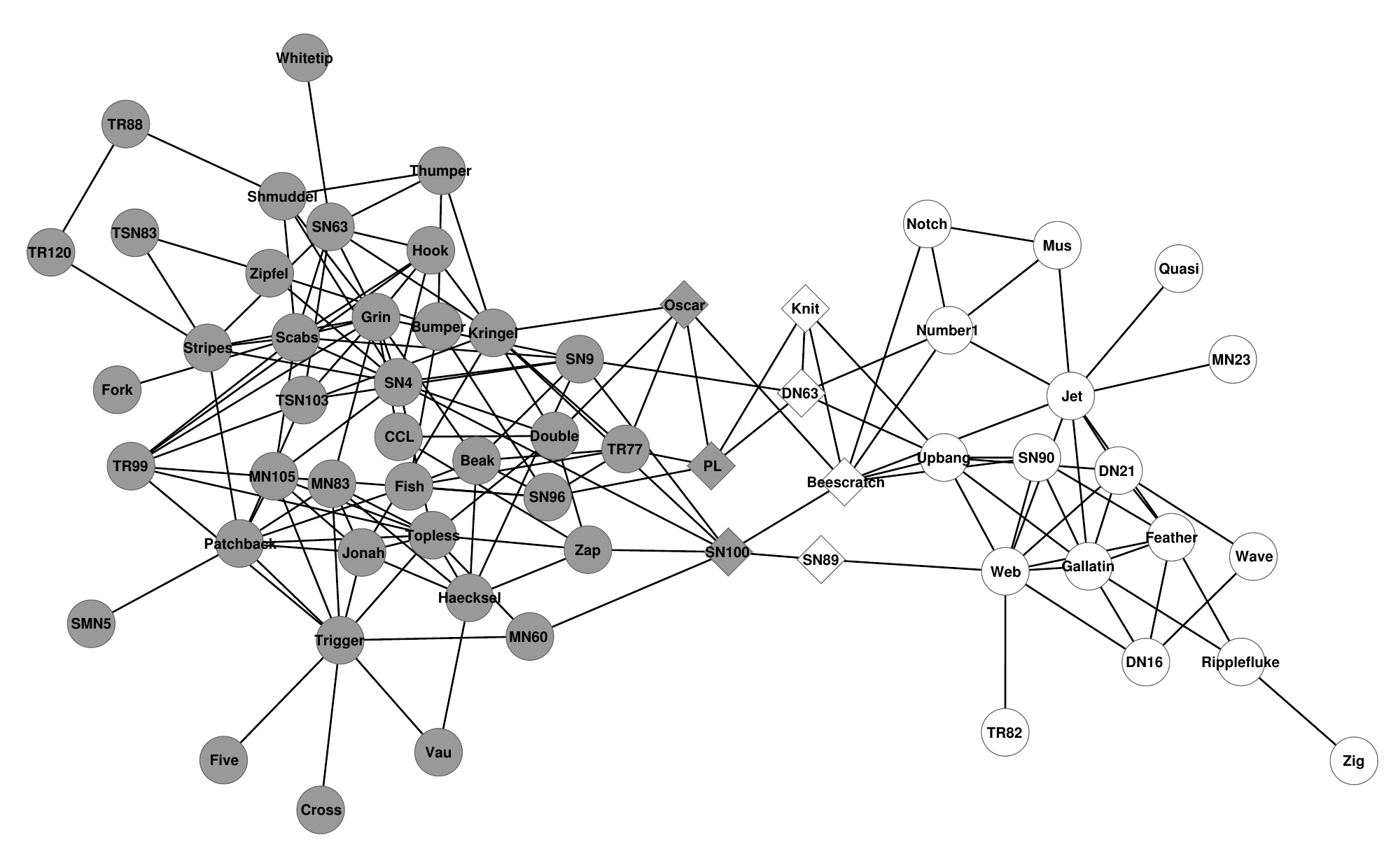}}
\caption{\label{fig:dolphin1}(a) Values of state vector number 61: the four levels indicate respectively nodes of community (white nodes), overlapping nodes (white diamond nodes), closer nodes of the other community(gray diamonds) and finally nodes of the other community (gray nodes).(b)Communities detected by our algorithm.}
\end{figure}

\section{Entropy of information}
\label{quinto}
In order to present the temporal results in a compact way, we computed the entropy $E$ of a configuration $S$, using the cumulative distribution over the non-normalized index,
\begin{equation}
 \begin{split}
  P_i^{(S)} &= \sum_i S_{ij},\\ 
  E^{(S)} &= - \sum_i P_i^{(S)} \log(P_i^{(S)}) - \sum_i (1-P_i^{(S)}) \log(1-P_i^{(S)}).
 \end{split}
\end{equation}

The entropy $E$ is maximal for the flat distribution, when each node knows only itself, and minimal (zero) what all the network has only one label (or has become just one star for the rewiring algorithm). If the population is evenly distributed among $n$ clusters, the entropy is $E=\log(n)$.  
 Let us to study the artificial complex network illustrated in~\figurename~\ref{f.lbl}. This network is composed by three levels with different probability to have a link in a region.
 
\begin{figure}[h!]
	\begin{center}
	\begin{tabular}{cc}
 		\includegraphics[width=0.45\columnwidth]{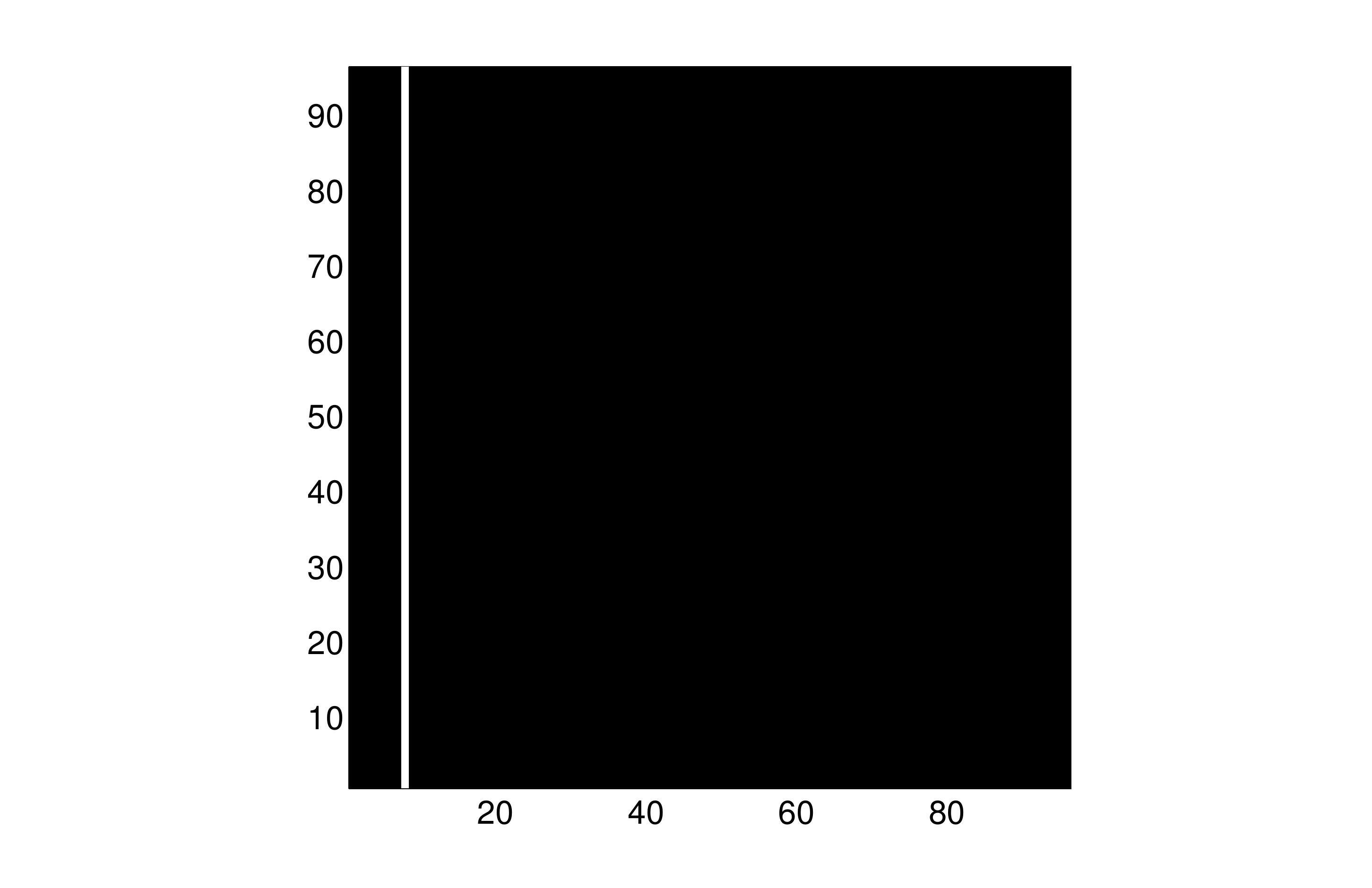}&
 		\includegraphics[width=0.45\columnwidth]{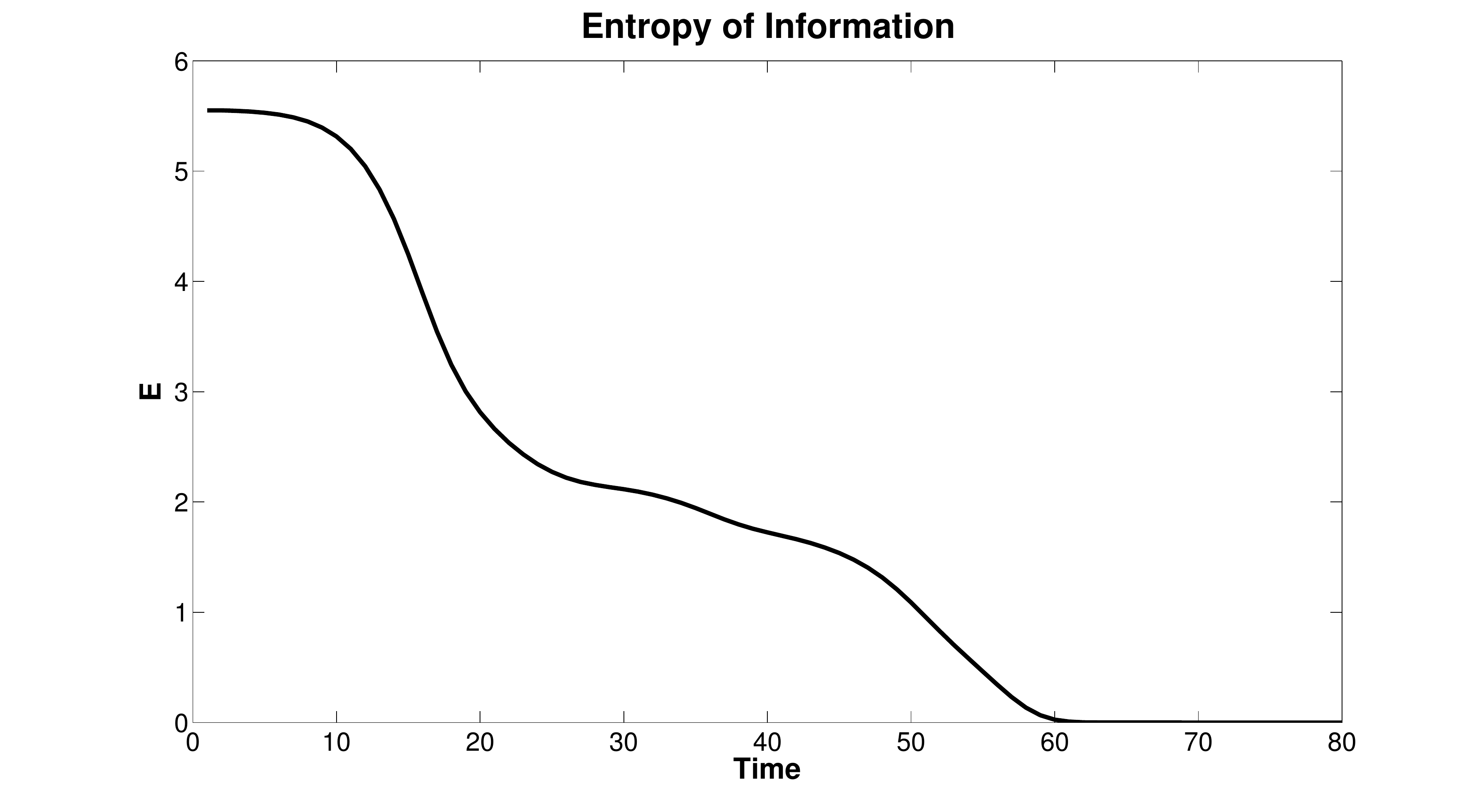}\\
	 (a)&(b)\\
	 &\\
	         \includegraphics[width=0.45\columnwidth]{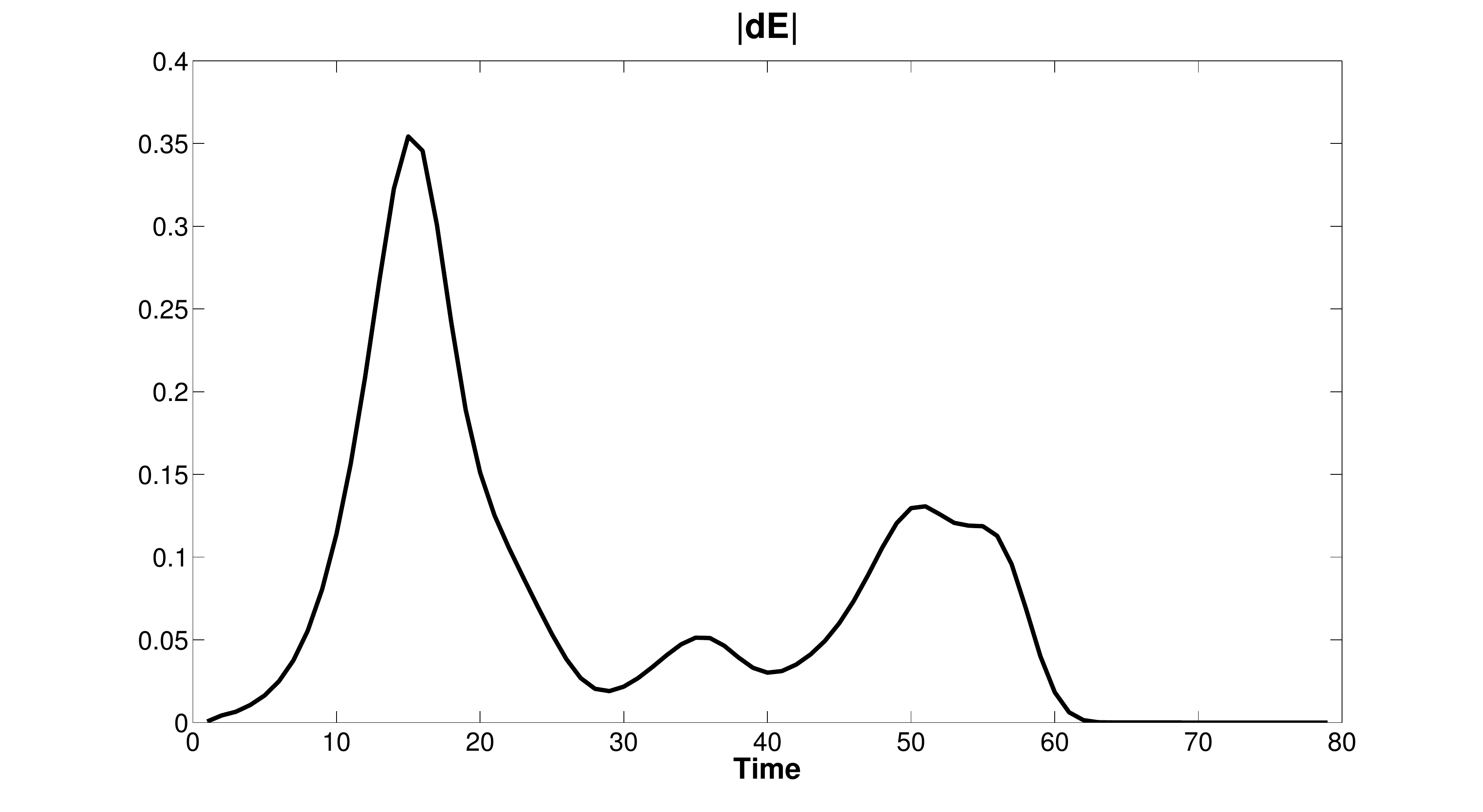}&
 		\includegraphics[width=0.45\columnwidth]{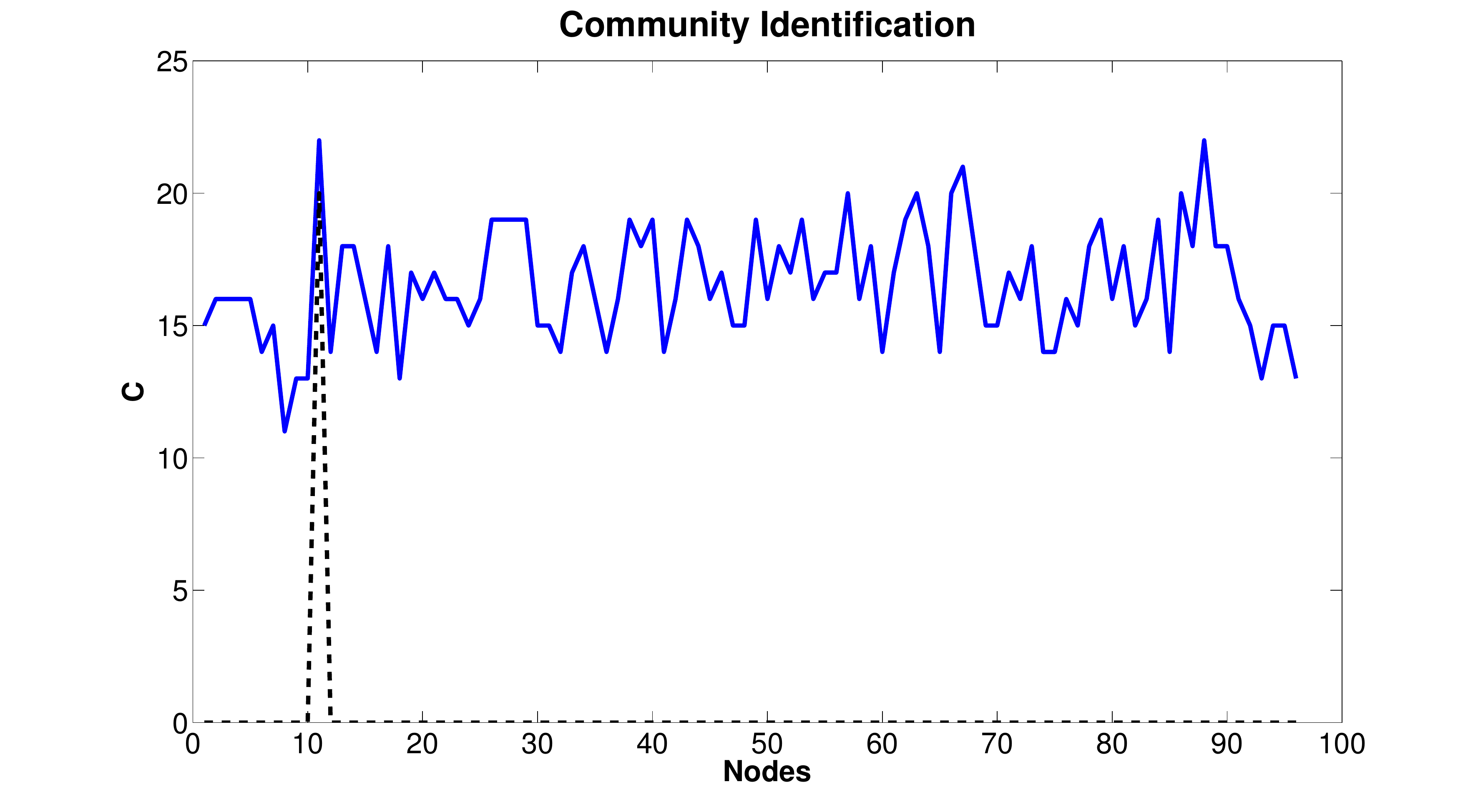}\\
 		
	(c)&(d)
	\end{tabular}
 	\end{center}
 	\caption{Referring to the artificial hierarchical network illustrated in~\figurename~\ref{f.lbl}: (a) the asymptotic configuration observed with our algorithm using $m=0.27$ and $\alpha=1.25$. (b) The Entropy of Information ($E$) vs Time: we can observe three different \emph{plateaux}. The final configuration, $E=0$, corresponds to the monocluster shown in (a); (c) plot of the first derivative of the Entropy which show the three \emph{plateu} with three different peaks. (d) We observe that the final community is identified by the higher connectivity.}
 		\label{fig:entropy1}
\end{figure}

 As we observed our algorithm is able to observe all levels of a hierarchical network. In~\figurename~\ref{fig:entropy1}(a) it is possible to identify the final level of the artificial network. The value of Entropy $E(t)$ can help us understand the structure of the network at \emph{priori}. In fact, different levels of a hierarchical structure are identified by the plateau as we can observe in~\figurename~\ref{fig:entropy1}(b). This result is emphasized by the entropy's first derivative where we can observe three distinct peaks~(\figurename~\ref{fig:entropy1}(c)). The final monocluster, using the adjacency matrix $A$ in the Eq.~\ref{ref:bagnongen}, is identified by the major hub in the  network~ (\figurename~\ref{fig:entropy1}(d)).
 
\begin{figure}[h!]
	\begin{center}
 		\includegraphics[width=1\columnwidth]{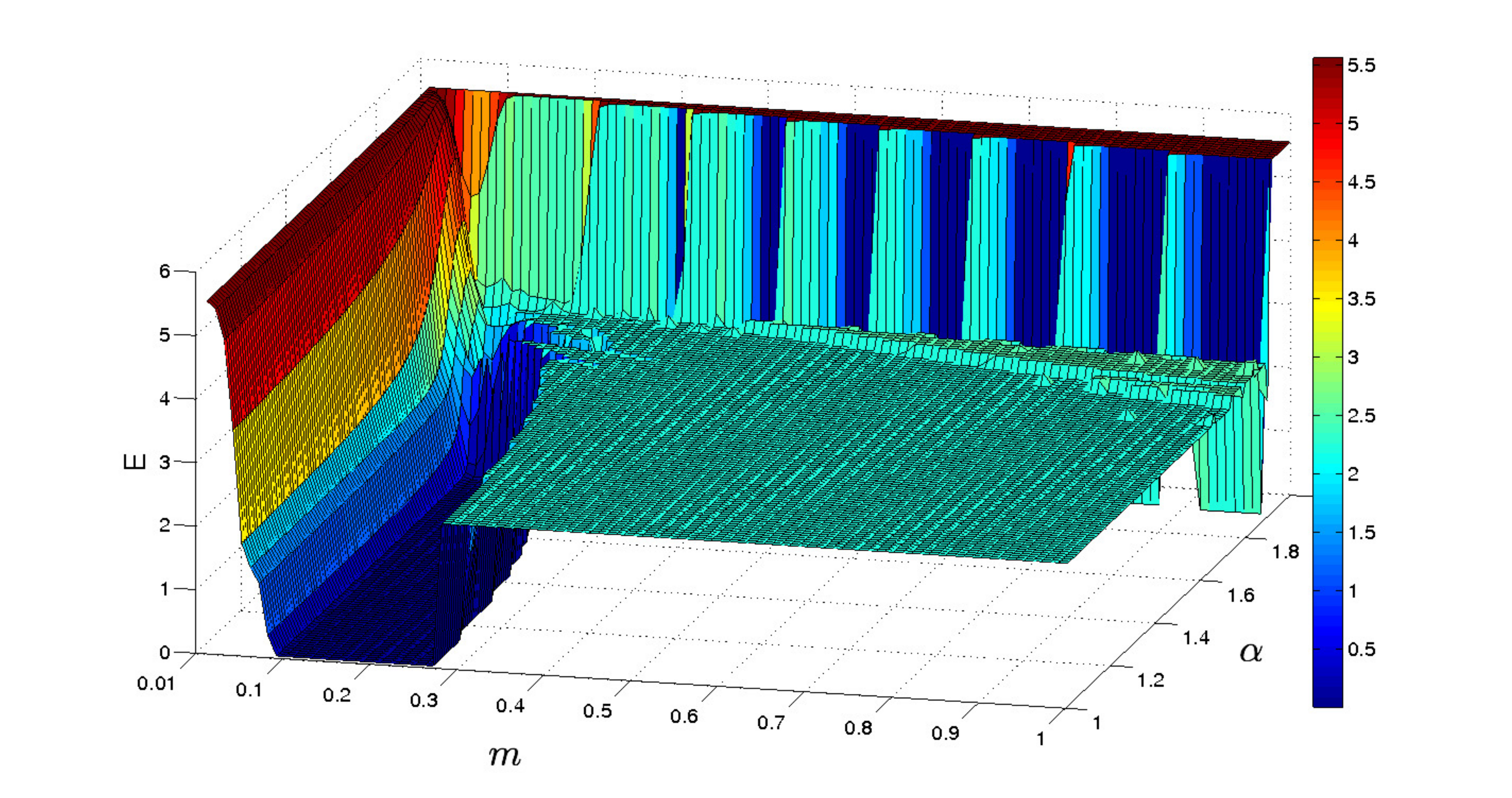}
 	\end{center}
 	\caption{The asympotic configuration of the entropy as function of the parameters  $m$ and $\alpha$. We can clearly observe many different surfaces in this 3D graph: the surfaces correspond to different asymptotic configurations}
 	\label{fig:entropy3d}
\end{figure}

 \begin{figure}[h!]
	\begin{center}
	\begin{tabular}{cc}
 		\includegraphics[width=0.45\columnwidth]{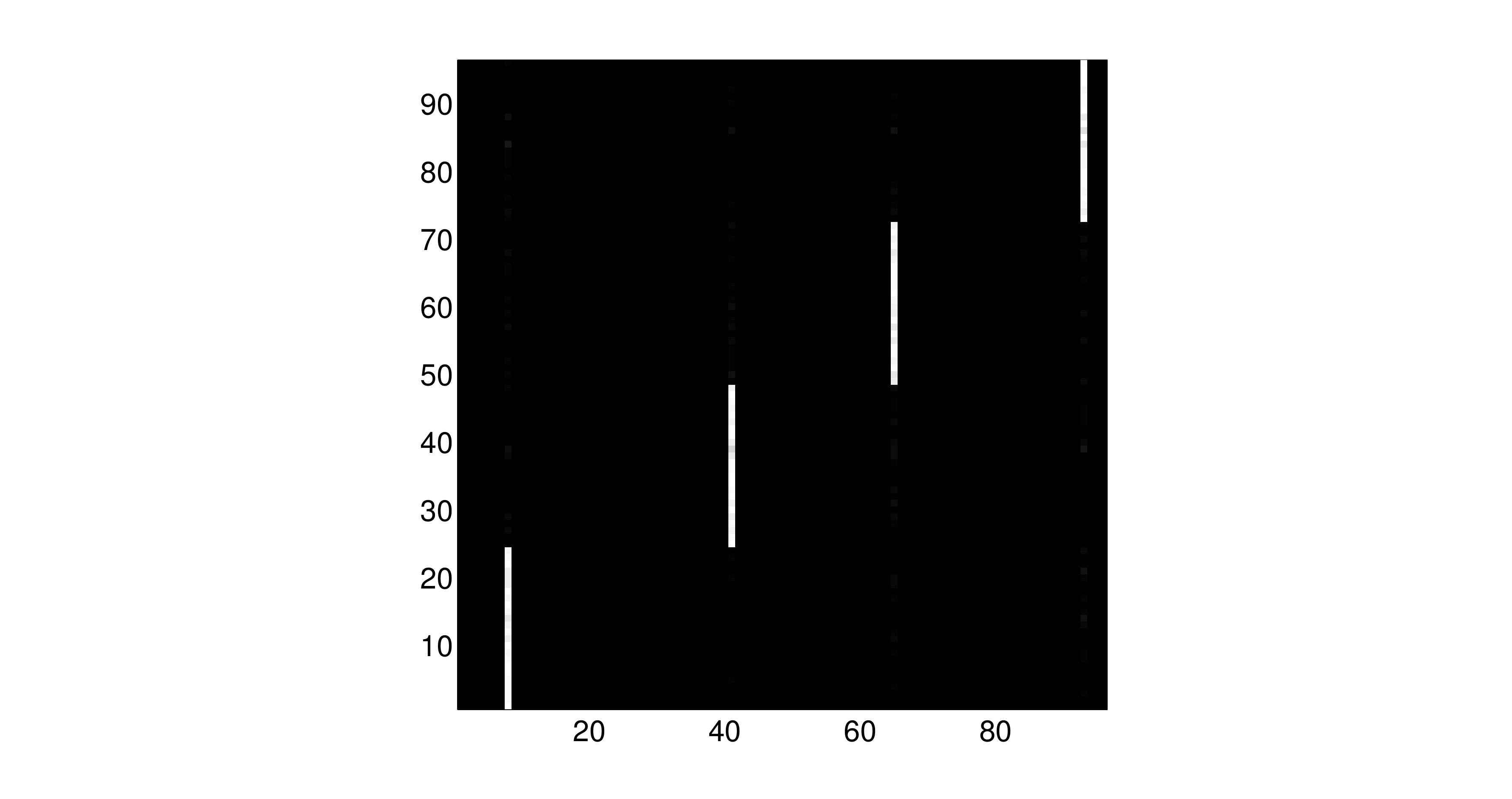}&
 		\includegraphics[width=0.45\columnwidth]{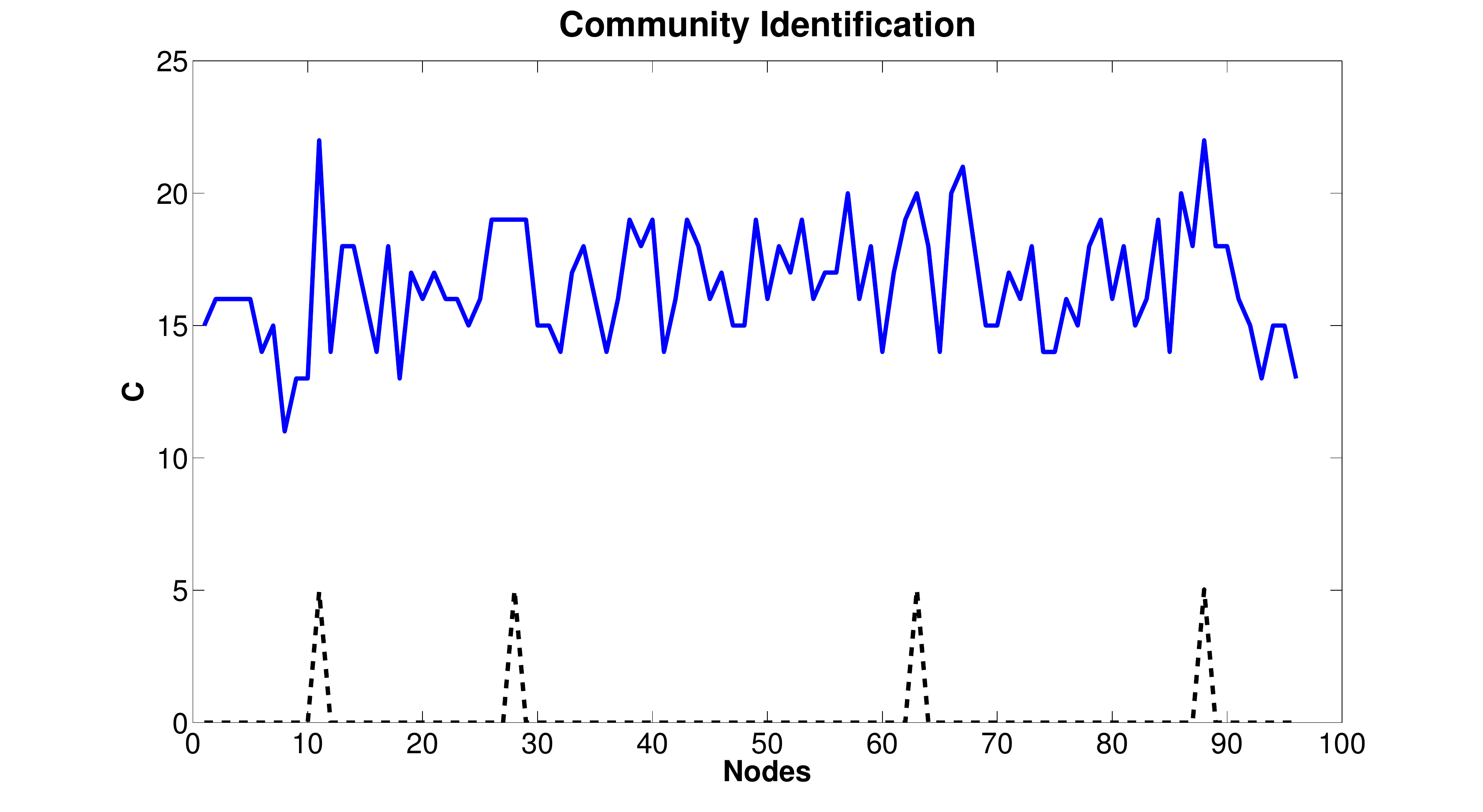}\\
	 (a)&(b)\\
	\end{tabular}
 	\end{center}
 	\caption{(a) Asymptotic configuration of the matrix $S$ using $m = 0.7$ and $\alpha = 1.4$ (this configuration corresponds to the large green surface in~\figurename~\ref{fig:entropy3d}). We observe that the asymptotic result corresponds to the middle layer (four communities) of the hierarchical structure of the network; (b) We have identified all the different levels in a hierarchical complex network changing the parameters $m$ and $\alpha$.}
 	\end{figure}

\section{Conclusion}
In this paper we have described an algorithm to identify the communities structures in a network from a local point of view.

The method is  based on pure information propagation where the nonlinear part of these method, responsible for the actual elaboration of information, is inspired by a chemical/ecological competition model~\cite{Nicosia09}.

There is not a unique definition of a community, so an exploratory algorithm, like the one that humans  have presumably developed during their evolution, should present different clustering for different values of the parameters, or for different iterations. 

In this implementation we adopted a frequency-based
approach and an unbounded memory at the level of nodes.
Unbounded memory means that the node’s state vector $S_i$
has not been limited and it could potentially reach a size
equal to the network size $N$. Despite this because of the
explanation and normalization  phases  are sufficient
to avoid this problem. Nevertheless it will be very important
to limit the computational resources of the node explicitly,
as suggested by Simon in 1955~\cite{Simon55}, so increasing both the
ecological plausibility of the model and the insights which drive
the algorithm design.

The results that we have obtained are promising. The method under investigation is not competitive with respect to others (see the review~\cite{Fortunato201075}), but it provides a natural ``scanning'' of the various clustering levels. Moreover, our method can be naturally applied to weighted graphs. We have demonstrated, through the definition of \emph{Entropy of Information}, our algorithm is efficient to discover  all cluster levels for general networks.

We believe that the local algorithm procedure will not only allow to us to study much larger networks but also to mimic single human behavior in social network trough specific and simple heuristics decision rules. The model parameters $m$ and $\alpha$ play a crucial role for the detection of communities. These results suggests how cognitive heuristics could be designed as those mechanism which allow humans to optimize those parameters in order to maximize the gathered information from the environment. Following this assumption the future works will investigate what kind of computational procedures could be used to mimic
this human behavior. We plan to investigate the consequences of bounded memory and computational resources of nodes, in particular in a dynamic environment.

\section*{Acknowledgments}
This work is financially supported by Recognition Project: RECOGNITION is a 7th Framework Programme project funded under the FET initiative.

\newpage
\section*{Bibliography}
\bibliography{biblio}

\begin{thebibliography}{23}
\providecommand{\natexlab}[1]{#1}
\providecommand{\url}[1]{\texttt{#1}}
\expandafter\ifx\csname urlstyle\endcsname\relax
  \providecommand{\doi}[1]{doi: #1}\else
  \providecommand{\doi}{doi: \begingroup \urlstyle{rm}\Url}\fi

\bibitem[Albert and Barabási(2002)]{Albert}
R.~Albert and A.~L. Barabási.
\newblock \emph{Rev. Mod. Phys.}, \penalty0 (74):\penalty0 47, 2002.

\bibitem[Blatt et~al.(1996)Blatt, Wiseman, and Domany]{domany}
M.~Blatt, S.~Wiseman, and E.~Domany.
\newblock Superparamagnetic clustering of data.
\newblock \emph{Phys. Rev. E}, \penalty0 (76):\penalty0 3251–3254, 1996.

\bibitem[Brandes et~al.(2006)Brandes, Delling, Gaertler, Gorke, Hoefer,
  Nikoloski, and Wagner]{clustering1}
U.~Brandes, D.~Delling, M.~Gaertler, R.~Gorke, M.~Hoefer, Z.~Nikoloski, and
  D.~Wagner.
\newblock {On finding graph clusterings with maximum modularity}.
\newblock In \emph{Proceedings of the 33rd International Workshop on
  Graph-Theoretical Concepts in Computer Science (WG’07)}, 2006.

\bibitem[de~Freitas et~al.(2000)de~Freitas, Lucena, da~Silva, and
  Hilhorst]{frei}
J.~E. de~Freitas, L.S. Lucena, L.R. da~Silva, and H.J. Hilhorst.
\newblock Critical behavior of a two-species reaction-diffusion problem.
\newblock \emph{Phys. Rev. E.}, \penalty0 (61):\penalty0 6330–6336, 2000.

\bibitem[Dongen(2009)]{MCL}
S.~Van Dongen.
\newblock Graph clustering via a discrete uncoupling process.
\newblock \emph{SIAM. J. Matrix Anal. and Appl.}, \penalty0 (30):\penalty0
  121--141, 2009.

\bibitem[Dorogovtesev and Mendes(2003)]{Mendes}
S.~N. Dorogovtesev and J.~F.~F. Mendes.
\newblock \emph{Evolution of Networks}.
\newblock Oxford University Press, Oxford, 2003.

\bibitem[Forster and Davis(1984)]{Forster84}
K.~I. Forster and C.~Davis.
\newblock Repetition priming and frequency attenuation.
\newblock \emph{Journ. Exp. Psyc.: Learning Memory and Cognition}, 10\penalty0
  (4), 1984.

\bibitem[Fortunato(2010)]{Fortunato201075}
Santo Fortunato.
\newblock Community detection in graphs.
\newblock \emph{Physics Reports}, 486\penalty0 (3-5):\penalty0 75 -- 174, 2010.
\newblock ISSN 0370-1573.
\newblock \doi{DOI: 10.1016/j.physrep.2009.11.002}.

\bibitem[Gigerenzer and Gaissmaier(2011)]{Gigerenzer2011}
G.~Gigerenzer and W.~Gaissmaier.
\newblock Heuristic decision making.
\newblock \emph{Ann. Rev. of Psyc.}, \penalty0 (62):\penalty0 451–482, 2011.

\bibitem[Gigerenzer and Goldstein(2002)]{Gigerenzer2002}
G.~Gigerenzer and G.~Goldstein.
\newblock Models of ecological rationality: The recognition heuristic.
\newblock \emph{Psyc. Rev.}, 109\penalty0 (1):\penalty0 75–90, 2002.

\bibitem[Lancichinetti et~al.(2009)Lancichinetti, Fortunato, and
  Kertész]{Lanc}
A.~Lancichinetti, S.~Fortunato, and J.~Kertész.
\newblock Detecting the overlapping and hierarchical community structure of
  complex networks.
\newblock \emph{New Journal of Physics}, 033015\penalty0 (11), 2009.

\bibitem[Lusseau et~al.(2003)Lusseau, Schneider, Boisseau, Haase, Slooten, and
  Dawson]{dolphin}
D.~Lusseau, K.~Schneider, O.~J. Boisseau, P.~Haase, E.~Slooten, and S.~M.
  Dawson.
\newblock \emph{Behavioral Ecology and Sociobiology}, \penalty0 (54):\penalty0
  396–405, 2003.

\bibitem[Murray(2002)]{murray2002}
J.D. Murray.
\newblock \emph{Mathematical biology}.
\newblock Number v. 1 in Interdisciplinary applied mathematics. Springer, 2002.
\newblock URL \url{http://books.google.it/books?id=aHoaKQEACAAJ}.

\bibitem[Newman(2004)]{communities}
M.E.J. Newman.
\newblock Detecting community structure in networks.
\newblock \emph{Europ. Phys. J. B}, \penalty0 (38):\penalty0 331--330, 2004.

\bibitem[Newman and Girvan(2004)]{Newman}
M.E.J. Newman and M.~Girvan.
\newblock Finding and evaluating community structure in networks.
\newblock \emph{Phys. Rev. E}, \penalty0 (69):\penalty0 026113, 2004.

\bibitem[Nicosia et~al.(2011)Nicosia, Bagnoli, and Latora]{Nicosia09}
V.~Nicosia, F.~Bagnoli, and V.~Latora.
\newblock Impact of network structure on a model of diffusion and competitive
  interaction.
\newblock \emph{EPL}, 94\penalty0 (68009), 2011.

\bibitem[Palla et~al.(2005)Palla, Derény, and Vickset]{Palla}
G.~Palla, I.~Derény, and T.~Vickset.
\newblock \emph{Nature}, 435:\penalty0 814, 2005.

\bibitem[Scott(2000)]{Scott}
J.~Scott.
\newblock \emph{Social Networks Analysis: A Handbook}.
\newblock Sage, London, 2nd, edition, 2000.

\bibitem[Simon(1955)]{Simon55}
H.A. Simon.
\newblock A behavioral model of rational choice.
\newblock \emph{The Quarterly Journal of Economics}, 69\penalty0 (1):\penalty0
  99--118, 1955.

\bibitem[Strogatz(2001)]{Strogatz}
S.H. Strogatz.
\newblock \emph{Nature(London)}, \penalty0 (410):\penalty0 268, 2001.

\bibitem[Tulving et~al.(1982)Tulving, Schacter, and Stark]{Tulving82}
E.~Tulving, D.~L. Schacter, and H.~A. Stark.
\newblock Priming effects in word fragment completion are independent of
  recognition memory.
\newblock \emph{Journ. Exp. Psyc.: Learning Memory and Cognition}, 8\penalty0
  (4), 1982.

\bibitem[Wasserman and Faust(1994)]{Waaserman}
S.~Wasserman and K.~Faust.
\newblock \emph{Social Networks Analysis}.
\newblock University Press, Cambridge, England, 1994.

\bibitem[Zachary(1977)]{Zachary}
W.~W. Zachary.
\newblock An information flow model for conflict and fission in small groups.
\newblock \emph{Journal of Anthropological Research}, \penalty0 (33):\penalty0
  452–473, 1977.

\end{thebibliography}
\bibliographystyle{plainnat}

\end{document}